\documentclass[11pt ]{article}
\usepackage[affil-it]{authblk}

\usepackage{graphicx}
\usepackage{amsmath, amsthm, amssymb}
\usepackage{subfigure}
\usepackage{comment}
\usepackage{bm}
\usepackage{epsfig,color}
\usepackage[sans]{dsfont}
\usepackage{verbatim}
\usepackage{amsfonts}
\usepackage{amsmath}
\usepackage{amssymb}
\usepackage{enumerate}

\newcommand{\E}{\mathbb{E}}

\newcommand{\tr}{\text{tr}}

\newcommand{\be}{\begin{equation}}
\newcommand{\ee}{\end{equation}}
\newcommand{\bea}{\begin{eqnarray}}
\newcommand{\eea}{\end{eqnarray}}
\newcommand{\bes}{\begin{equation*}}
\newcommand{\ees}{\end{equation*}}
\newcommand{\beas}{\begin{eqnarray*}}
\newcommand{\eeas}{\end{eqnarray*}}

\renewcommand{\H}{\mathcal{H}}

\def\B{\mathcal{B}}

\def\ido{\mathrm{I}}
\def\i{\mathrm{id} }

\def\tr{\mbox{tr}}

\def\E{\mathcal{E}}

\def\tr{\mathrm{tr}}

\newcommand{\id}{\rm{id}}

\newtheorem*{thm*}{Theorem}

\newtheorem*{lem*}{Lemma}

\newtheorem*{lipschitzLem*}{Lemma \ref{lipschitz}}
\newtheorem*{lipschitzCubeLem*}{Lemma \ref{lipschitzCube}}
\newtheorem*{pgmNearlyOptimalThm*}{Theorem \ref{pgmNearlyOptimal}}
\begin{document}

\title{ Designing Quantum Information Processing via Structural Physical Approximation  }

\author{Joonwoo Bae$^{1,2}\footnote{bae.joonwoo@gmail.com}$ }

\affil{
$^{1}$Department of Applied Mathematics, Hanyang University (ERICA), 55 Hanyangdaehak-ro, Ansan, Gyeonggi-do, 426-791, Korea, and  \\
$^{2}$ Freiburg Institute for Advanced Studies (FRIAS), Albert-Ludwigs University of Freiburg, Albertstrasse 19, 79104 Freiburg, Germany \\ }

\maketitle

\begin{abstract}
In quantum information processing it may be possible to have efficient computation and secure communication beyond the limitations of classical systems. In a fundamental point of view, however, evolution of quantum systems by the laws of quantum mechanics is more restrictive than classical systems, identified to a specific form of dynamics, that is, unitary transformations and, consequently, positive and completely positive maps to subsystems. This also characterizes classes of disallowed transformations on quantum systems, among which positive but not completely maps are of particular interest as they characterize entangled states, a general resource in quantum information processing. Structural physical approximation offers a systematic way of approximating those non-physical maps, positive but not completely positive maps, with quantum channels. Since it has been proposed as a method of detecting entangled states, it has stimulated fundamental problems on classifications of positive maps and the structure of Hermitian operators and quantum states, as well as on quantum measurement such as quantum design in quantum information theory. It has developed efficient and feasible methods of directly detecting entangled states in practice, for which proof-of-principle experimental demonstrations have also been performed with photonic qubit states. Here, we present a comprehensive review on quantum information processing with structural physical approximations and the related progress. The review mainly focuses on properties of structural physical approximations and their applications toward practical information applications. 
\end{abstract}

\newpage
\section{Introduction}

Information processing with quantum systems may provide advantages over the currently existing limitations on the computational and information capabilities of classical systems. Applying quantum systems to computational tasks, the information processing is governed by the laws of quantum mechanics, wherein quantum resources are generated during the evolution such as superposition, entanglement, and quantum interference. It turns out that, in this way, the the prime factorization problem can be efficiently solved with quantum systems and their evolution \cite{Shor:1999aa}. Searching a target in a unsorted database can be formulated as the amplitude amplification algorithm that also leads to a quadratic speedup with respect to the classical counterpart \cite{Grover:1996:FQM:237814.237866}, which is also optimal \cite{Zalka:1999aa}. 

Entangled states, that is, quantum correlations that have no classical counterpart \cite{schrdinger1935, Werner:1989aa, RevModPhys.81.865}, are generally a resource for quantum information processing. Highly entangled states endowed with local measurements can perform computational tasks \cite{Raussendorf:2001aa}. When entangled states are shared by legitimate parties, maximally entangled states can be distilled \cite{Bennett:1996aa} and entanglement swapping can be performed \cite{Zukowski:1993aa, Briegel:1998aa}, or they can be converted by local measurement to secret correlations \cite{Curty:2004aa, Acin:2005aa} so that they can be applied to quantum communication protocols. Entanglement states can also establish secret key for information-theoretically secure communication, see for instance Ref. \cite{PhysRevLett.83.4200}.

In the fundamental point of view, there are actually the postulates of quantum theory behind all that quantum information processing is distinguished from the classical counterparts. It is worth mentioning that among physical theories, a unique feature of quantum theory is its formalism that they are given in the form of axioms on physical entities, quantum states, dynamics, and measurement. Quantum dynamics is postulated to be a unitary transformation by which the aforementioned computational advantages can be achieved. Entanglement existing in multipartite quantum systems allows it possible to have non-classical effects in quantum communication, for instance, super-activation effects \cite{Smith:2008aa, Palazuelos:2012aa}. Note that these do not generally correspond to measurable quantities, in contrast to classical systems in which physical entities are identified by measurable quantities. 

Then, postulates of quantum theory, at the same time, also characterize disallowed dynamics, that is, non-unitary evolution often related to impossible tasks in quantum information processing. For instance, a pair of non-orthogonal states together cannot be transformed by quantum dynamics to mutually orthogonal ones. This can be restated as the impossibility of perfectly distinguishing non-orthogonal quantum states, that is closely related to other no-go theorems such as the no-cloning and the no-signaling principle \cite{Wootters:1982aa, Gisin:1998aa, Bruss:1998aa, Bae:2006aa, Chiribella:2006aa}. Disallowed dynamics is then directly linked to practical applications: for instance, the aforementioned impossibility can be directly applied to secure quantum communication, e.g., \cite{Bennett:1992aa}. 

Note that when dynamics of quantum systems is governed by a unitary transformation, the description of subsystem's dynamics is characterized by positive and completely positive (CP) maps over quantum states \cite{JdeP, Jamiokowski:1972aa, Choi:1975aa}, see also for instance the open quantum systems in Ref. \cite{book:breuer}. Positive but non-CP maps, which are thus disallowed in quantum theory, precisely identify the set of all entangled states in the sense that these maps transform all quantum states but separable ones to non-positive operators that cannot be interpreted as quantum states. Conversely, for an entangled state, there exists a positive but non-CP map that detects the state \cite{Woronowicz:1976aa, Horodecki:1996aa}. All these reiterate the significance of disallowed positive maps that can detect entangled states for quantum information processing to lead to the quantum advantages. 

Structural physical approximation (SPA), initially proposed in Ref. \cite{Horodecki:2003ab} to devise approximating to nonlinear functionals on quantum states, then offers a systematic way of  constructing a physical process that approximate positive but non-CP maps. Once SPA is applied to the positive maps, the resulting approximate map which thus corresponds to a quantum channel is henceforth no longer able to detect entangled states. Then, one can naturally ask how the aforementioned quantum advantages are affected by SPA with a view taken from entanglement theory. 

There has been remarkable progress in both theoretical and implementation sides of SPA and entanglement theory. The conjecture in Ref. \cite{SPAconjecture} addressed that SPA leads to separable states, and has been an intriguing problem in both technical and experimental aspects. While being supported by numerous examples \cite{SPAconjecture, Chruscinski:2009aa, Chruscinski:2010aa, Chruscinski:2011aa, Zwolak:2013aa, Zwolak:2014aa, Augusiak:2014aa}, finally it has been disproved by counterexamples \cite{Ha:2012aa, Stormer:2013aa, Chruscinski:2014ab, Hansen:2015ab}. On the experimental side, SPA has been exploited to realize quantum channels that approximate disallowed dynamics such as transpose and partial transpose \cite{Lim:2011aa, PTexpBae, Lim:2012aa}. Apart from the fundamental interest, these may be building blocks to entanglement detection and also for quantum information applications in general. It turns out that SPA can introduce the so-called quantum design \cite{Lim:2012aa, KalevBae, Graydon:2016ac}, a specific form of POVMs, that is of both fundamental and practical interest in quantum information theory. Recently, an excellent review has been presented with a focus on the mathematical structure of SPA and the conjecture \cite{Shultz:2016aa}. 

We here present a comprehensive review on SPA and the conjecture with a view taken from quantum information applications. We mainly focus on the interplay between SPA and entangled states and its applications to processing and realizing quantum information tasks. When SPA leads to an entanglement-breaking quantum channel, its implementation is hugely simplified to an experimentally feasible scheme, that only performs measurement and preparation of quantum states. Then, quantum measurement involved in SPA has a particular structure called quantum design, both of fundamental and practical interest in quantum information theory. Nonetheless, positive maps are not always transformed to entanglement-breaking channels by SPA.

The paper is organized as follows. In Sec. \ref{section:pre}, we summarize quantum theory and introduce terminologies and notations to be used throughout. In Sec. \ref{section:entanglement}, we review the entanglement theory briefly about characterization and detection of entangled states. In Sec. \ref{section:spatheory}, we introduce SPA to positive maps and provide its properties. In Sec. \ref{section:spaexp}, we review experimental progress in implementation of the approximate transpose and the approximate partial transpose. In Sec. \ref{section:applications}, we present recent progress in applications of SPA to entanglement detection. In Sec. \ref{section:conclusion}, we conclude with a summary on the progress in SPA and address open questions.



\setcounter{footnote}{0}
\section{States, Dynamics, and Measurement}\label{section:pre}
\markboth{\sc States, Dynamics, and Measurement}{}

Let us begin with summarizing the formalism and collecting terminologies and notations to be used throughout. As it is mentioned, quantum theory is formalized with axioms on physical entities such as states, dynamics, and measurement. The formalism can be described with operators in Hilbert space. Let $\H_{d}^{(A)}$ denote a $d$-dimensional Hilbert space of a quantum system $A$. If the dimension is clear from the context, the subscript is omitted and it is written as $\H$. Let $\B(\H)$ denote the set of bounded operators in Hilbert space $\H$.

{\bf States.} In quantum theory, a state is described by a bounded, linear, and non-negative operator on a Hilbert space. To have the interpretation to probabilities, operators describing quantum states are of unit-trace. Let $S(\H)$ denote the set of quantum states on Hilbert space $\H$,
\bea
S(\H) = \{ \rho \in \B(\H)~ : \tr\rho = 1,~\rho\geq 0~ \}. \nonumber
\eea
For multipartite systems, a state is described by bounded, non-negative and unit-trace operators on $\H\otimes \cdots \otimes \H $. 

In the space $S(\H)$, pure states correspond to extremal operators as they cannot be expressed by a convex combination of other states. A pure state thus corresponds to a rank-one operator. Equivalently, a state $\rho$ is pure if and only if $\tr[\rho^2] = 1$. Otherwise, a state is called a mixed state that is not of rank-one, and also $\tr[\rho^2] <1$. 

Mixed states can be described in equivalent and alternative ways in the following. The first is that mixed states are given when $a ~priori$ knowledge is lacking in state preparation. Suppose that a party Alice prepares state $\{ \rho_i\}_{i=1}^n$ according to probabilities $\{ p_i\}_{i=1}^n$, respectively, and then sends it to the other, Bob. Then, on average, Bob's state is described as $\rho_B = \sum_i p_i \rho_i$. Note that preparation of mixed states is not unique. 

Mixed states are also given as a marginal of entangled states. For a state of system $\rho^{(S)}$, there always exists a purification, which means a pure state of system and environment $E$, $\rho^{(SE)} \in S( \H\otimes \H^{(E)})$ such that $\tr_E \rho^{(SE)} = \rho^{(S)}$. Purifications are equivalent up to local unitary transformations. Suppose that system and environment are in the following purification,
\bea
\rho^{(SE)} = | \psi^{(SE)} \rangle \langle \psi^{(SE)} |,~~\mathrm{where} ~ |\psi^{(SE)} \rangle = \sum_{i=1}^N \sqrt{ p_i }  | \psi_{i}^{(S)} \rangle  | \psi_{i}^{(E)} \rangle. \label{eq:purification}
\eea
Then, discarding environment, the system state is necessarily given by a mixture of pure states as $\tr_E \rho^{(SE)} = \sum_{i=1}^N  p_i |\psi_{i}^{(S)} \rangle \langle \psi_{i}^{(S)} |$.  In other words, system's being in a mixed state arises from entanglement between system and environment.

To describe entangled states, say for bipartite system of two parties Alice and Bob $\rho^{(AB)} \in S(\H^{(A)} \otimes \H^{(B)})$, one has to introduce local operations and and classical communication (LOCC), that actually characterize separable states in an operational way. Suppose that Alice and Bob can prepare quantum states using local operations $\rho^{(A)} \otimes \rho^{(B)}$, and they can also communicate each other via classical means. This allows them to prepare a number of product states probabilistically. Those quantum states that can be prepared in this way are called separable states and can be written in the following form
\bea
\rho_{\mathrm{sep}}^{(AB)}  = \sum_{i} p_i \rho_{i}^{(A)}\otimes  \rho_{i}^{(B)}. \label{eq:sep-state}
\eea
Then, bipartite quantum states that are not in the form in Eq. (\ref{eq:sep-state}) are called entangled states.

{\bf Measurement.} Measurement on quantum systems produces outcomes in a probabilistic way. The measurement postulate dictates the mapping from quantum states to probabilities via positive-operator-valued-measures (POVMs), which are given as
\bea
M_{i} \geq 0 ~~\mathrm{for} ~~i =,1\cdots,n ~~ \mathrm{such~ that}~ ~\sum_{i=1}^n M_i = \ido. \nonumber
\eea
That is, POVMs are a positive resolution of the identity operator.

In experimental realization, each POVM element $M_i$ correspond to a description of a detector. Suppose that there are $n$ detectors for measurement on state $\rho$. A complete measurement means that for any state $\rho$, one of the $n$ detectors must show a detection event, {\it click}. Then, for instance, let the $j$th detector is described by POVM $M_j$. From the postulate of quantum theory, the probability of having a detection event on $M_j$ is given by 
\bea
p (M_j | \rho) =\tr[M_j \rho], \label{eq:born}
\eea
which is called the Born rule. In fact, the Born rule constructs the unique probability measure \cite{gleason}. As the relation in Eq. (\ref{eq:born}) shows conditional probabilities, it holds that
\bea
\forall \rho \in S(\H),~~\sum_{j} p(M_j | \rho) =1. \nonumber 
\eea
This implies that $\sum_j M_j = \ido$, the completeness condition for POVMs.

In general, POVM elements can be implemented via the so-called Naimark's dilation theorem. It shows that one can implement POVMs in general via orthogonal measurement on additional systems, in a similar vein of the existence of purifications for quantum states in Eq. (\ref{eq:purification}). To be precise, it states that any POVM element can be implemented with an additional ancilla system and orthogonal measurement on the ancilla: for POVM $M_i$, there exist environment $\rho_{i}^{(E)}$, unitary transformation $U^{(SE)}$, and orthogonal measurement $\{ P_{i}^{(E)}\}$ such that 
\bea
\tr [\rho M_i] = \tr [ U^{(SE)} ( \rho \otimes \rho_{i}^{(E)} ) U^{(SE) {\dagger}} ~P_{i}^{(E)} ]. \label{eq:naimark1}
\eea
This shows that for a given system $\rho$, measurement on POVM $\{ M_i\}$ can be equivalently implemented by orthogonal measurement on ancillas after making dilation on the system. The right-hand-side in Eq. (\ref{eq:naimark1}) can be written as
\bea
\tr[\rho M_i],~\mathrm{where}~ M_i =  \tr_E [  \rho_{i}^{(E)}  U^{(SE) {\dagger}} P_{i}^{(E)}  U^{(SE)}]. \nonumber
\eea
This shows a method of devising POVMs in experimental implementation.

{\bf Dynamics.} There are equivalent and alternative descriptions to quantum dynamics. Let us first present the description with isometry. Suppose that a quantum system evolves for time $0$ to $t$ while interacting with environment. Recall that the overall dynamics must be unitary, denoted by $U_{t}^{(SE)}$, as it is postulated. We also assume that an environment state $\rho^{(E)}$ is initially decoupled from system. Then, a quantum operation can be described by the dynamics reduced to system as follows, 
\bea
\rho~ \mapsto~ \rho_t = \tr_E ~U_{t}^{(SE)} \rho \otimes \rho^{(E)} U_{t}^{(SE) \dagger}.  \label{eq:stinespring}
\eea
Fixing the environment state as $\rho^{(E)} = | 0\rangle_E \langle 0| $, one can find the isometry,
\bea
V_t = \tr_{E} U_{t}^{(SE)} (\ido_S \otimes |0\rangle_E),~\mathrm{so~that}~ \rho ~\mapsto~ \rho_t = V_t \rho V_{t}^{\dagger} \nonumber
\eea
where it holds that $V_{t}^{\dagger} V_t= \ido_S$. Note that in the description above, called Stinespring dilation  \cite{Stinespring:1955aa}, it is essential that system and environment are initially in a completely factorized form. Otherwise, the map in Eq. (\ref{eq:stinespring}) does not give a legitimate description on dynamics of quantum systems. 

The above can be equivalently described in the Kraus representation \cite{krausrep}. A set of operators $\{ K_i \}_{i=1}^n$, which are not positive in general, are called Kraus operators if they satisfy $\sum_{i=1}^n K_{i}^{\dagger} K_i =\ido$. Then, dynamics of a quantum state can be described by a set of Kraus operators such that
\bea
\rho~ \mapsto~ \rho_t = \sum_i K_i \rho K_{i}^{\dagger}. \label{eq:kraus}
\eea
In Eq. (\ref{eq:stinespring}), fixing $\rho^{(E)} = | 0\rangle_E \langle 0|$ and having denoted orthonormal basis $\{ |j\rangle_E \}_j$ in environment, one can relate the Stinespring dilation with Kraus operators as follows,
\bea
K_{j}  = _{E}\langle j | U^{(SE)} |0 \rangle_E,~~\mathrm{scuh ~that}~~\sum_{j} K_{j}^{\dagger} K_j = \ido_S. \nonumber 
\eea
This shows that once environment is found in state $| j\rangle_E\langle j |$, it implies that the system has evolved under Kraus operator $K_j$. That is, the resulting state is given by $(\rho_{t})_j =  K_j \rho K_{j}^{\dagger}/p_j$ with $p_j = \tr[\rho K_{j}^{\dagger} K_j]$. If it is not informed which state the environment is in, the system is described as a probabilistic mixture, $\rho_t = \sum_j p_j (\rho_t)_j$, as it is shown in Eq. (\ref{eq:kraus}). 

After all, quantum operations can be characterized by linear maps over quantum states. A linear map $\Lambda : \B(\H) \rightarrow \B(\H)$ is called positive, denoted by $\Lambda \geq 0$, if it maps a positive operator to another positive one, i.e.,
\bea 
\Lambda \geq 0 ~ ~ \iff ~ ~\Lambda[\rho] \geq 0,~~ \forall \rho \in S(\H). \nonumber 
\eea
The definition can be generalized to $k$-positivity: $\Lambda$ is $k$-positive, $\i_k \otimes \Lambda\geq 0$, that is, 
\bea 
(\i_k \otimes \Lambda ) [\rho] \geq 0, ~~ \forall \rho \in S( \H_{k}^{(E)}\otimes \H),\nonumber
\eea
where $\H_{k}^{(E)}$ denotes $k$ dimensional environment and $\i_k$ denote the identity map in the $k$-dimensional space. Then, a linear map corresponding to a quantum operation must be positive on Hilbert space of system, i.e. a positive map, and positive also on Hilbert space of system and arbitrarily extended environment. A map $\Lambda$ is called completely positive (CP) if it is $k$-positive for all $k\geq1$. A positive and CP map can be implemented as a physical process. Conversely, a physical process can be described by a positive and CP maps in general. 

Note that in the above, for the dimension of ancilla systems, it suffices to consider dimension $k$ up to the system dimension. That is, a linear map $\Lambda$ describes a quantum operation if $\Lambda \geq 0$ and $\i \otimes \Lambda \geq 0 $ where $\i$ is the identity map on Hilbert space of environment whose dimension is as large as the system. We also call a quantum operation $\Lambda$ trace-preserving if it holds that $\tr[\Lambda [ \rho ] ] = 1$ for all $\rho\in S(\H)$. A trace-preserving quantum operation is then referred to as a quantum channel.


\setcounter{footnote}{0}
\section{Entanglement Theory}\label{section:entanglement}
\markboth{\sc Entanglement Theory}{}

In this section, we summarize characterization and quantification of entangled states. We also discuss feasible methods of detecting entangled states.

\subsection{Characterization and quantification}  
\label{subsection:characterization}


We first recall that separable states are those quantum states that can be prepared by LOCC. They can be written in general as follows, 
\bea
\sigma_{\mathrm{sep}} = \sum_i p_i \rho_{i}^{(A)} \otimes \rho_{i}^{(B)},~~\mathrm{for}~\rho_{i}^{(A)}\in S(\H_A) ~\mathrm{and}~ \rho_{i}^{(B)} \in S(\H_B).  \label{eq:separable}
\eea
Separable states can be obtained by locally preparing $\rho_{i}^{(A)}$ and $\rho_{i}^{(B)}$ and communicating the probabilities $\{ p_i\}$. An important property is the convexity. Separable states form a convex set: a probabilistic mixture of separable states is also separable. In mathematical terms, separable states are the dual to positive maps in a operator space. That is, those positive operators that remain positive under all positive maps are characterized as separable states. We write separable states as, denoted by $\mathrm{SEP}$
\bea
\mathrm{SEP} = \{ \sigma\in S(\H^{(A)}  \otimes \H^{(B)} )~\| ~(\i \otimes  \Lambda) [\sigma] \geq 0,~\forall \Lambda\geq0  \} \nonumber
\eea
for all positive maps $\Lambda$.

\begin{figure}
\begin{center}  
\includegraphics[width= 10cm]{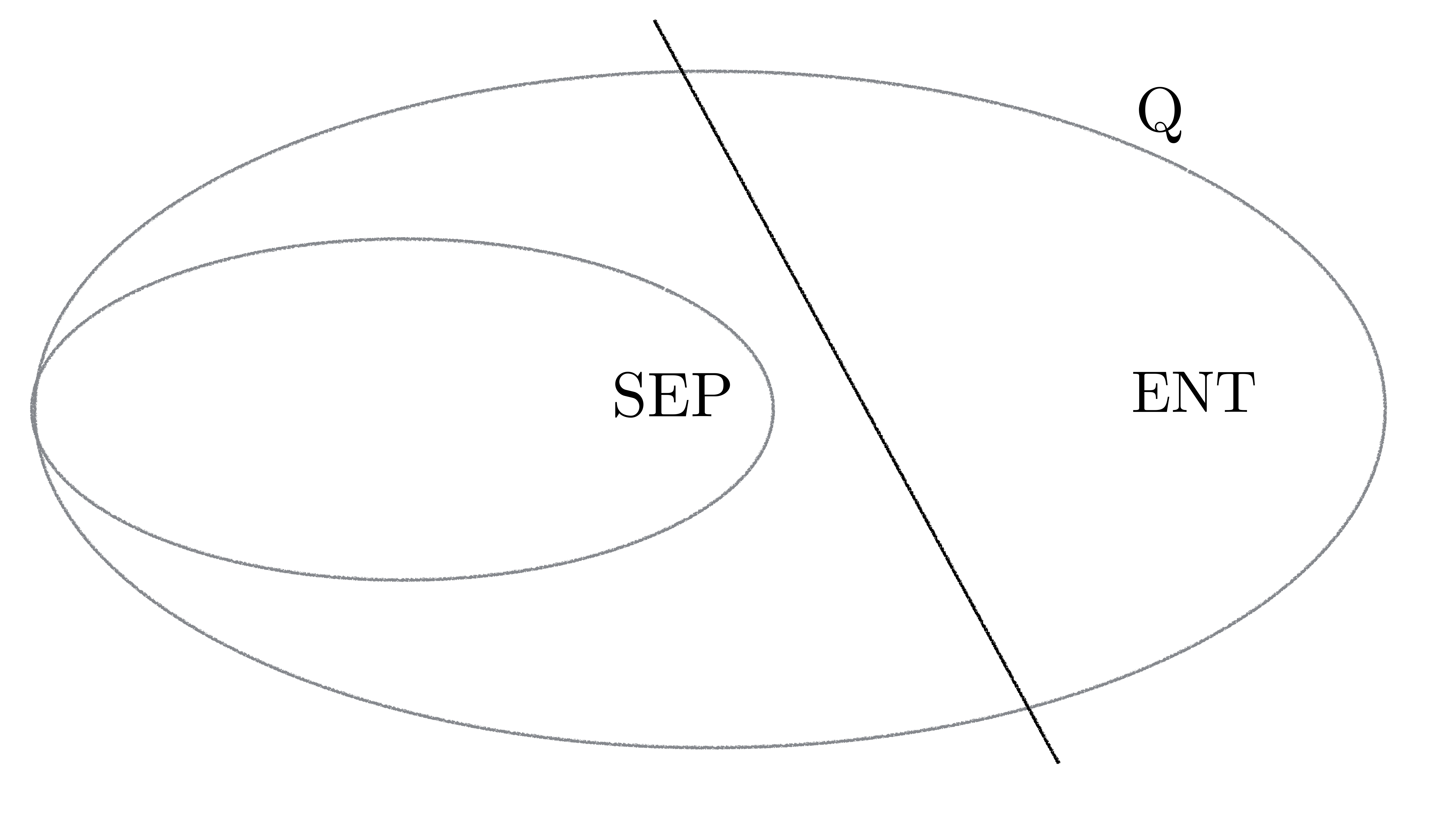}
\caption{The set of separable states is convex. An entangled state can be distinguished by a hyperplane from separable states. } 
\label{fig:structure}
\end{center}
\end{figure}

Entangled states are those quantum states that cannot be prepared by LOCC, not possible to be written in the form of Eq. (\ref{eq:separable}). They do not form a convex set: a mixture of entangled states can be a separable state. Note also that the set of bipartite quantum states is the dual to the CP maps $\i\otimes \Lambda \geq 0$, i.e. 
\bea
\mathrm{Q} = \{ \rho \in S(\H^{(A)}  \otimes \H^{(B)} )~\| ~(\i \otimes \Lambda) [\rho] \geq 0,~\forall ~\i \otimes \Lambda \geq 0  \}. \nonumber
\eea
Then, entangled states denoted by $\mathrm{ ENT}$ corresponds to the complement to separable states, $\mathrm{ENT} =Q \setminus \mathrm{SEP}$. This shows that positive but not CP maps give the characterization as the dual to entangled states. In fact, all entangled states can be detected by positive but non-CP maps \cite{Horodecki:1996aa}.  

When it is found given systems are in an entangled states, the next is quantification of entanglement. We recall that LOCC is the operational task that can prepare only separable states but entangled ones, i.e., LOCC does not generate entanglement. An entanglement measure $E$ therefore has to fulfull the following constraints 
\bea
\mathrm{ i)} &&  E(\rho_{\mathrm{ent}})>0~~ \mathrm{for ~all ~entangled ~states}~\rho_{\mathrm{ent}},  \nonumber \\
\mathrm{ii)} &&  E(\sigma_{\mathrm{sep}}) = 0 ~~\mathrm{for~all~  separable ~states} ~\sigma_{\mathrm{sep}}. \label{eq:Econdition}
\eea
Since LOCC does not increase entanglement, for states $\rho$ and $\sigma$ the measure $E$ satisfies the following property,
\bea
\rho ~\rightarrow_{\mathrm{LOCC}} ~ \sigma ~~ \Rightarrow ~~ E (\rho) \geq E(\sigma), \label{eq:more-ent}
\eea
where $\rightarrow_\mathrm{LOCC}$ denotes an LOCC protocol transforming state $\rho$ to $\sigma$. It is clear that $\sigma$ is not more entangled than $\rho$. Then, it follows that
\bea
\rho~ \rightarrow_{\mathrm{LOCC}}~\sigma~\mathrm{and}~ \sigma~ \rightarrow_{\mathrm{LOCC}}~\rho ~\iff~ E(\rho) = E(\sigma), \nonumber
\eea
meaning that $\rho$ and $\sigma$ are equally entangled, or that they are equivalent up to local unitaries: there exist local unitaries $U = U^{(A)}\otimes U^{(B)} $ such that $U\rho U^{\dagger} = \sigma$.

The relation in Eq. (\ref{eq:more-ent}) immediately shows that LOCC gives an order relation among quantum states. In fact, the set of bipartite states is totally ordered under LOCC, i.e. for any pair of states $\rho$ and $\sigma$, either $\rho\rightarrow_{\mathrm{LOCC}} \sigma$ or $\sigma\rightarrow_{\mathrm{LOCC}} \rho$ holds true. For states $\rho$, $\sigma$, and $\gamma$, we also have 
\bea
\rho\rightarrow_{\mathrm{LOCC}}\sigma ~\mathrm{and}~\sigma\rightarrow_{\mathrm{LOCC}}\gamma ~\Longrightarrow ~ \rho\rightarrow_{\mathrm{LOCC}}\gamma. \nonumber
\eea
Moreover, there is a unique root state in the order structure up to local unitaries such that all other states can be prepared by LOCC. The root state must be more entangled than any other states, for which it is called maximally entangled, and is given by in $S(\H_d\otimes \H_d)$
\bea
|\phi_{d}^{+}\rangle  = \frac{1}{\sqrt{d}} (|11\rangle + \cdots + |dd\rangle)\label{eq:maxent}
\eea
We remark that the maximally entangled state can be identified only with the order relation with LOCC. A function of multipartite quantum states is called an entanglement monotone \cite{Vidal:2000aa} if it satisfies the conditions in Eqs. (\ref{eq:Econdition}) and (\ref{eq:more-ent}), see also computable entanglement measures in Refs. \cite{Wootters:1998aa, Vollbrecht:2001aa, Vollbrecht:2001aa, Terhal:2000aa, Mintert:2004aa, Mintert:2005aa, Mintert:2005ab, Mintert:2007ab, Bae:2009aa}

\subsection{Positive maps and entanglement witnesses}  
\label{subsection:entdetection}

Entangled states can be characterized by positive maps or, equivalently, entanglement witnesses (EWs). Both can detect entangled states. Entanglement detection is of both theoretical and practical importance as the characterization of entangled or separable states is highly non-trivial and entanglement is generally a useful resource for quantum information processing. When positive maps are attempted to apply to decide if given states are entangled or separable, one has to first completely identify given quantum states beforehand, with quantum state tomography. On the other hand, by applying EWs, entanglement can be detected even before learning given states with tomography.

In what follows, we show details of two aforementioned approaches of entanglement detection. We here restrict the consideration to single-copy level measurement, that is feasible with current technologies. Note that there are more efficient approaches that applies collective measurement on milti-copies, e.g. \cite{Augusiak:2008aa, Mintert:2007aa}. Collective measurement is in general experimentally challenging as  quantum memory is required to store quantum states for a while.

We first recall that positive but non-CP maps give the characterization of entangled states, vice versa. The condition that a map $\Lambda$ is positive but not CP can be rephrased by the followings,
\bea
&& i)~ (\i \otimes \Lambda ) [\sigma_{\mathrm{sep}}] \geq 0,~~\forall~\sigma_{\mathrm{sep}} \in \mathrm{SEP},~~\mathrm{and}~~ \nonumber\\
&& ii) ~ \exists ~\rho\in \mathrm{ENT}~~ \mathrm{such~that} ~ (\i \otimes \Lambda) [\rho] \ngeq 0. \label{eq:pncp}
\eea
Equivalently, a state $\rho$ is entangled if and only if there exists a positive but non-CP map $\Lambda$, 
\bea
\exists \Lambda\geq 0, ~\i \otimes \Lambda \ngeq 0,~~\mathrm{such~that}~~ (\i\otimes \Lambda)[\rho] \ngeq 0. \nonumber
\eea
Entangled states can be identified by positive but not-CP maps. Note that, however, it has been a longstanding open problem in the context of operator algebra to have a complete characterization of positive but non-CP maps. Alternatively, it is also one of major challenging problems in quantum information theory to characterize separable states. This is referred to as the separability problem, which turns out to be in the NP-hard class \cite{Gurvits:2004:CCQ:1039323.1039332}. 

Despite the fact that the decision problem itself is intractable, there have been fruitful directions with known examples of positive but non-CP maps. The first instance is the transpose operation, denoted by $T$,
\bea 
T: \rho = \sum_{ij} \rho_{ij} | i\rangle \langle j| ~\mapsto~ T[\rho] = \sum_{ij} \rho_{ij} | j \rangle \langle i |. \nonumber
\eea
The operation $\i \otimes T$ is called partial transpose and written as $\Gamma$. For state $\rho \in S(\H^{(A)} \otimes \H^{(B)})$, if it is found that $(\i \otimes T) [\rho] \ngeq 0$, one can conclude that the state is entangled \cite{Peres:1996aa}. The converse does not hold true in general: that is, there exist entangled states that remain positive under the partial transpose \cite{Horodecki:1999aa}.

\begin{figure}
\begin{center}  
\includegraphics[width= 10cm]{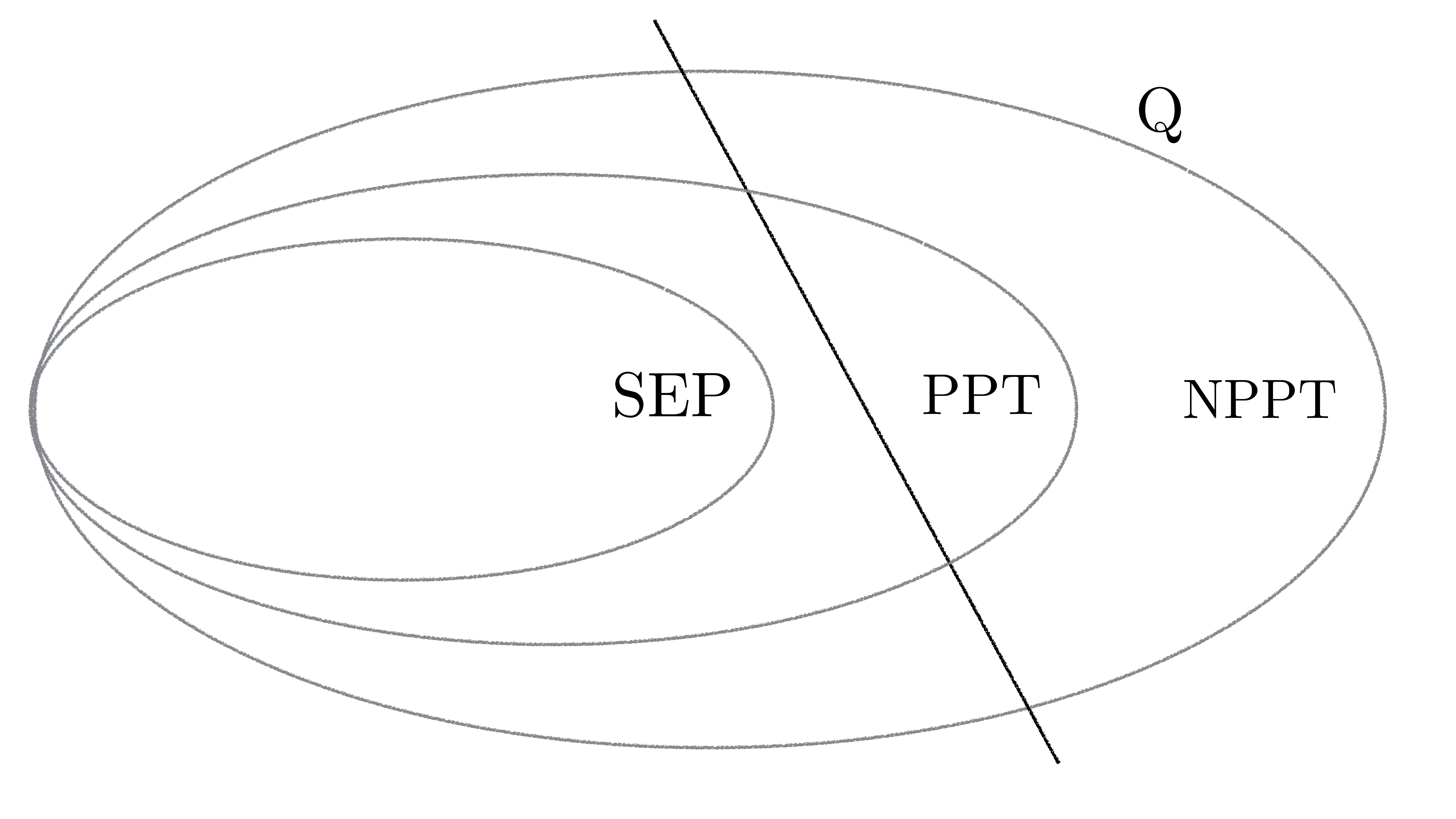}
\caption{ PPT states also form a convex set, and contains the set of separable states. NPPT states are not convex.} 
\label{fig:pstructure}
\end{center}
\end{figure}

In fact, the partial transpose gives a simple criteria of identifying useful quantum states. Let us write those quantum states remaining (non-)positive after the partial transpose by (N)PPT, as follows,
\bea
\mathrm{PPT}  & = &  \{ \rho \in S(\H^{(A)}\otimes \H^{(B)})~\|~\rho^{\Gamma} \geq 0  \} \nonumber \\ 
\mathrm{NPPT} & = &  \{ \rho \in S(\H^{(A)} \otimes \H^{(B)}) ~ \| ~ \rho^{\Gamma} \ngeq 0  \} \nonumber.
\eea 
It is clear that $\mathrm{SEP} \subset \mathrm{PPT}$. Note also that for $\mathrm{dim}(\H^{(A)}) =2~\mathrm{and}~\mathrm{dim}(\H^{(B)}) \in \{2,3\}$ we have that 
$\mathrm{SEP} = \mathrm{PPT}$ \cite{Peres:1996aa, Horodecki:1996aa}. There are entangled states which remain positive under the partial transpose, which are called PPT entangled states (PPTES). No entanglement can be distilled from PPTES. Note that for $\mathrm{dim}(\H^{(A)}) =2~\mathrm{and}~\mathrm{dim}(\H^{(B)}) \in \{2,3\}$, a positive map $\Lambda$ has a canonical form that
\bea
\Lambda = \Lambda_{1}  + T\circ  \Lambda_{2} \label{eq:canform}
\eea
for some CP maps $\Lambda_{1} $ and $\Lambda_{2}$. In general, a positive map that can be written as the form in Eq. (\ref{eq:canform}) is called decomposable.

An instance of decomposable maps is the reduction map, $\Lambda_R : S(\H) \rightarrow S(\H)$,
\bea
\Lambda_R [\rho ]  = \frac{1}{d-1} (\ido -\rho). \nonumber
\eea
Then, the map $\i \otimes \Lambda_R$ is particularly useful as a distillability criteria \cite{Horodecki:1999ab}. If a state is detected by the reduction map, i.e., $[\i \otimes \Lambda_R] (\rho) \ngeq 0$, then it is not only entangled but also distillable. Note that decomposable maps can detect only NPPT states.  

Positive maps that are not in the form in Eq. (\ref{eq:canform}) are called indecomposble, and can detect PPTES. A well-known example is Choi's map, $\Lambda_C : S(\H_3) \rightarrow S(\H_3)$,
\bea
\Lambda_C [\rho] = \frac{1}{2} ( -\rho + \sum_{i=0}^2 \rho_{ii} (2 |i\rangle \langle i| + |i-1 \rangle \langle i-1 |  ) ~) \label{eq:choimap}
\eea
where the $|i-1\rangle = |i-1~\mathrm{mod}~3~\rangle$. That is, the map $\i \otimes \Lambda_C$ is not positive for some PPTES.

EWs that can detect entanglement of unknown states, i.e., even before verification of quantum states, can be constructed as follows. Let us restate the condition in Eq. (\ref{eq:pncp}): a state $\rho$ is entangled if and only if there exists a positive but non-CP map $\Lambda$ such that $[\i \otimes \Lambda](\rho) \ngeq 0$. This means that there exists a projector $Q \geq 0$ such that 
\bea
\tr [ ( \i \otimes \Lambda ) [\rho] ~Q ]  < 0\nonumber
\eea
that is, $Q$ is one of the projectors onto the subspace in which $( \i \otimes \Lambda ) [\rho] $ contain negative eigenvalues. Note that there exists a dual map $\Lambda^{\dagger}$ such that the following holds true
\bea
  \tr [ ( \i \otimes \Lambda ) [\rho] ~ Q ] = \tr [ \rho ~ ( \i \otimes \Lambda^{\dagger} ) [ Q ] ].  \label{eq:dual} 
\eea
Let us write by $W =  ( \i \otimes \Lambda^{\dagger} ) [ Q ]$ so that the right-hand-side can be written as $\tr[\rho W]$. Note also that the operator $W$ is Hermitian, $W = W^{\dagger}$.

In Eq. (\ref{eq:dual}), suppose that $\rho$ is separable,. Then, the left-hand-side is positive since $\Lambda$ is positive and $Q\geq 0$. Thus, we have $\tr[\rho W] \geq 0$ for all separable states $\rho$. When $\rho$ is entangled, then there exists a map $\Lambda$ and $Q\geq 0$ such that the left-hand-side is negative. Then, we have $\tr[ \rho W] < 0$ for some entangled states $\rho$. To summarize, we have
\bea
&& \tr[\sigma_{\mathrm{sep}} W] \geq 0~~\mathrm{for~all~separable~states}~\sigma_{\mathrm{sep}} \nonumber \\
&& \tr[\rho_{\mathrm{ent}} W]  < 0 ~~\mathrm{for~some~entangled~states~} \rho_{\mathrm{ent}} \label{eq:EW}
\eea
These Hermitian operators are called EWs since they distinguish some entangled states from all separable states \cite{Terhal:2000ab, Guhne:2009aa, Chruscinski:2014aa}.

\setcounter{footnote}{0}

\section{Structural Physical Approximation and Quantum Channels}\label{section:spatheory}
\markboth{\sc Structural Physical Approximations and Quantum Channels}{}

Positive but non-CP maps that characterize entangled states do not correspond to a physical process, since they may take positive operators representing quantum states to non-positive ones that have no way to be interpreted as quantum states or probabilities. In Ref. \cite{Horodecki:2003ab}, a systematic way of transforming those non-CP maps to CP maps has been proposed and called structural physical approximations (SPA). In other words, SPA finds a quantum channel that approximates a positive but non-CP map.  

\subsection{Structural physical approximation to positive maps}


Let $\Lambda: \B(\H^{(A)}) \rightarrow \B( \H^{(B)})$ denote a positive but non-CP map which does not correspond to a physical process. SPA to the map is given by,
\bea
\widetilde{ \Lambda } &=& (1-p^{*}) \Lambda + p^{*} D_{A\rightarrow B} \nonumber\\
&& \mathrm{with}~ D_{A\rightarrow B}(\cdot) = \frac{ 1}{d_B} \ido ~ ~\mathrm{such ~that~~} \i \otimes \widetilde{ \Lambda} \geq 0 \label{eq:spa}
\eea
where $p^{*}$ denotes the minimum $p$ that $\widetilde{\Lambda}$ is CP, and $d_B$ the dimension of Hilbert space $\H^{(B)}$. Note also that the depolarization map is denoted by $D_{A\rightarrow B} (\cdot) = \ido / d_B$ with identity operator $\ido$. The SPAed map $\widetilde{ \Lambda}$ corresponds to a physical operation which can be implemented in experiment. Since the identity operator $\ido$ is of full rank, there exists non-trivial $p^{*} \in (0,1)$ such that $\widetilde{\Lambda}$ is CP. The term, "structural", comes from the fact that the depolarization map $D_{A\rightarrow B}$ is admixed, which does not modify the structure of original map $\Lambda$ \cite{Horodecki:2003ab}. 

The construction of SPA can be generalized by considering more possibilities of CP maps in the place of depolarization map in Eq. (\ref{eq:spa}) \cite{Augusiak:2011aa}. With a full-rank and normalized operator $K$, a generalization of SPA is given by
\bea
\widetilde{ \Lambda }_K = (1-p_{K}^{*}) \Lambda + p_{K}^{*} \E_K,~\mathrm{such ~that~} \i \otimes \widetilde{ \Lambda}_K \geq 0, \nonumber
\eea  
where $\E_K (\cdot) = K$. It is noteworthy that, to have a non-trivial $p_{K}^{*}\in(0,1)$, the operator $K$ should be of full-rank.

In general, SPA can be applied to non-positive maps $\i \otimes \Lambda$ which can detect entangled states as follows,
\bea
\widetilde{\i \otimes \Lambda} &=& (1-p^{*}) \i \otimes \Lambda +  { p^{*} }D_A \otimes D_B,  \label{eq:spa-ent} \\
&& \mathrm{where}~ p^* = \frac{\lambda d_{A}^3 d_B}{1 + \lambda d_{A}^3 d_B} ~\mathrm{and} ~\lambda = - \min_{Q>0} \tr[Q(\i\otimes \Lambda)[P_{d_A}^+] ]  \nonumber
\eea
such that $\widetilde{\i \otimes \Lambda}\geq 0$ is CP, where $D_{i}(\cdot) = \ido_i /d_i$ for $i=A,B$ denote the complete depolarization channel \cite{SPAconjecture}. Note that $p^*$ is found as a minimal $p$ that the resulting map is CP. 

The SPAed map in Eq. (\ref{eq:spa-ent}) can be applied to detecting entangled states as follows. Since $\rho\in S(\H^{(A)}\otimes \H^{(B)} )$ is entangled if $(\i\otimes \Lambda)[\rho]\ngeq 0$ for some positive map $\Lambda\geq 0$, we have that $\rho$ is entangled if
\bea 
\widetilde{ \i \otimes \Lambda} [\rho] \ngeq \frac{\lambda d_A d_B}{ \lambda d_{A}^3 d_B +1}. \label{entspa}
\eea
The difference is that, whereas the map $ \i \otimes \Lambda$ is not a physical process, SPAed map $\widetilde{ \i \otimes \Lambda}$ corresponds to a quantum channel that can be experimentally realized. Therefore, the condition in Eq. (\ref{entspa}) can be applied to entanglement detection in practice by incorporating to estimation of minimum eigenvalues. The scheme for entanglement detection has been proposed in Ref. \cite{Horodecki:2002aa} together with the spectrum estimation in Ref. \cite{Keyl:2001aa}. The proposal is remarkable in that it directly applies positive maps to entanglement detection and also it provides an alternative approach to EWs.  

Moreover, the SPAed map in Eq. (\ref{eq:spa-ent}) can be implemented by an LOCC protocol \cite{Alves:2003aa}. For the purpose, the inversion map has been introduced, $\Theta = -\i:S(\H^{(A)}) \rightarrow S(\H^{(A)})$ and its SPA can be constructed as follows,
\bea
\widetilde{\Theta} = \frac{1}{d_{A}^2 -1 } \Theta + \frac{  d_{A}^2}{d_{A}^2 -1 } D_{A\rightarrow A},\label{eq:SPAtheta}
\eea
which is CP. Note that the inversion map is not even positive. Then, for a positive map $\Lambda : \B (\H^{(A)}) \rightarrow \B(\H^{(B)})$, the LOCC scheme to realize SPA to the map $\i\otimes\Lambda$, see Eq. (\ref{eq:spa-ent}), can be found by the decomposition in the following,
\bea
\widetilde{\i \otimes \Lambda} &=& (1 - q)~ \i \otimes \widetilde{\Lambda} + q~ \widetilde{\Theta} \otimes D_{A\rightarrow B} ~~\mathrm{with}~ q = \frac{ \lambda d_{A}^3 d_B -  \lambda d_{A} d_B  }{1+ \lambda d_{A}^3 d_B}~~~~~~~ \label{eq:spaLOCC}  
\eea
where $P_{d_A}^+$ denotes the projection onto the $d_A$-dimensional maximally entangled state, see Eq. (\ref{eq:maxent}). The obtained decomposition shows that the map $\widetilde{\i \otimes \Lambda}$ can be implemented by performing local operations $ \i \otimes \widetilde{\Lambda}$ and $\widetilde{\Theta} \otimes D_{A\rightarrow B}$ with probabilities $(1-q)$ and $q$, respectively. The LOCC scheme is useful when two parties far in distance implement the SPAed map and apply it to detecting entangled states.

Recall that positive maps and quantum states are closely related. One may observe how the relation between entanglement and positive maps evolves by SPA, by which positive maps are no longer non-CP and thus cannot detect entangled states. This has been elaborated and addressed as a conjecture that SPAed maps of optimal positive maps would characterize separable states \cite{SPAconjecture}. For cases where the conjecture holds true, there is a huge simplification in implementation, namely that SPAed maps can be implemented via a measure-and-prepare protocol without entangled resources. This can also be applied to local operations in the SPAed map in Eq. (\ref{eq:spaLOCC}). 

\subsection{Quantum channels and entanglement}

In this subsection, we collect machineries to discuss relations between SPA and entanglement. The characterizations to positive maps, EWs, entangled states, separable states, and quantum channels, that have been discussed so far, can be viewed in a coherent way with the so-called Choi-Jamio\l{}kowksi (CJ) isomorphisim \cite{JdeP, Jamiokowski:1972aa, Choi:1975aa}. It shows the one-to-one correspondence between the set of linear maps $\B(\H^{(A)}) \rightarrow \B(\H^{(B)})$ and bipartite operators in $\B(\H^{(A)}\otimes \H^{(B)})$. By the isomorphism, a map $\Lambda : \B(\H^{(A)}) \rightarrow \B(\H^{(A)})$ and a bipartite operator $\chi \in \B(\H^{(A)} \otimes \H^{(B)})$ are related as
\bea
&&\Lambda  (\cdot)  = d_A \tr_A [ \chi ~ ((\cdot)^{T} \otimes \ido)],~~\mathrm{and}~~ \label{eq:CJ1}\\
&&\chi =  (\i \otimes \Lambda ) [P_{d_A}^{+}]  \label{eq:CJ2}
\eea
where $P_{d_A}^{+} = |\phi_{d_A}^{+}\rangle \langle \phi_{d_A}^{+} | $ and $|\phi_{d_A}^{+}\rangle$ is the maximally entangled state in $S(\H_A \otimes \H_A)$ in Eq. (\ref{eq:maxent}). We note that Eq. (\ref{eq:CJ1}) shows that the linear map $\Lambda$ can be constructed from a given bipartite operator $\chi$ and conversely, Eq. (\ref{eq:CJ2}) that the bipartite operator $\chi$ can be obtained from a linear map $\Lambda$. Throughout, let $\chi_{\Lambda}$ denote the CJ operator for a map $\Lambda$. 

The CJ isomorphism is a useful tool in quantum information theory. It offers a unified view to the structure of positive maps and bipartite operators. Let us summarize the main results along the line, in the following.

\begin{figure}
\begin{center}  
\includegraphics[width= 12cm]{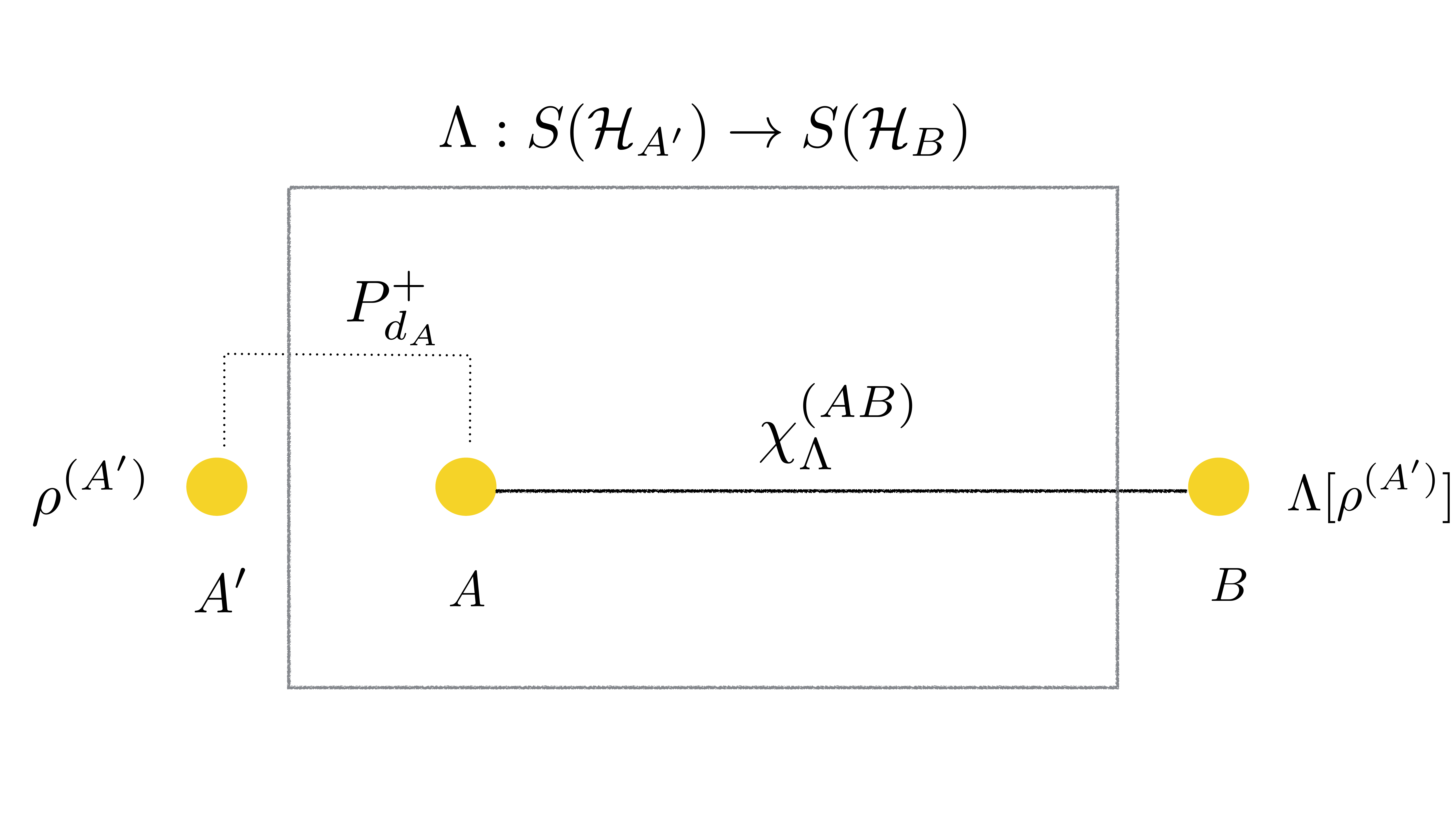}
\caption{ The CJ isomorphism shows that a quantum channel $\Lambda:S(\H^{(A')})\rightarrow S(\H^{(B)})$ can be implemented equivalently by the quantum teleportation protocol that exploits a shared state $\chi_{\Lambda}^{(AB)} = (\i \otimes \Lambda ) [P_{d_A}^{+} ]$, where $d_A = d_{A'}$. A state $\rho^{(A')} \in S(\H^{(A')})$ is teleported to the other site at $B$, where the resulting state appears as $\Lambda [ \rho^{(A')}]$ due to the shared state $\chi_{\Lambda}^{(AB)}$ that gives the complete characterization of the channel.} 
\label{fig:CJ}
\end{center}
\end{figure}

\begin{enumerate}
\item A linear map $\Lambda : S(\H^{(A)})\rightarrow S(\H^{(B)})$ is a quantum channel, i.e., CP and trace-preserving map, if and only if its CJ operator $\chi_{\Lambda}$ is a quantum state in $S(\H_A \otimes \H_B)$, i.e., positive and of unit-trace. This means that all quantum states in $S(\H^{(A)} \otimes \H^{(B)})$ are obtained by sending the maximally entangled state $|\phi_{d_A}^{+} \rangle$ via a quantum channel $\i \otimes \Lambda$. Conversely, any quantum channel $\Lambda$ can be characterized by a quantum state $\chi$ as it is shown in Eq. (\ref{eq:CJ1}). 

To see this explicitly, one can rewrite Eq. (\ref{eq:CJ1}) as follows \cite{Cirac:2001aa}
\bea
\Lambda  [ (\cdot) ] =   d_{A}^2 ~ \tr_{A'A} [ ( \cdot)_{A'}^{T} \otimes  \chi^{(AB)} ~ ( | \phi_{d_A}^{+}\rangle_{A'A} \langle \phi_{d_A}^{+ }  |) ]  \label{eq:CJ3}
\eea
where we have assumed that $d_{A'} = d_A$. This can be interpreted as quantum teleportation of quantum state at location $A'$ via shared entangled state $\chi^{(AB)}$ between $A$ and $B$, see also Fig. \ref{fig:CJ}. Note that the factor $d_{A}^{2}$ amounts to the success probability of measurement on the basis $P_{d_{A}}^+$ \footnote{Bell-states used in two-qubit quantum teleportation can be generalized to high-dimensions with Weyl operators, $\mathcal{W}_{m,n} = \sum_{k} e^{ \frac{2\pi i}{d}kn } | k \rangle \langle k+m |$ for $m,n=0,\cdots, d-1$, so that $| \Phi_{m,n} \rangle =(\i \otimes \mathcal{W}_{m,n} ) | \Phi_{d_A}^{+}\rangle$. }.

\item A linear map $\Lambda$ is positive but non-CP if and only if its CJ operator $\chi_{\Lambda}$ is an EW. That is, an EW can be obtained from a positive but non-CP map in general, as
\bea
W_{\Lambda} = (\i \otimes \Lambda ) [P_{d_A}^{+} ]. \label{eq:WL}
\eea
EWs are called (in-)decomposable if positive maps $\Lambda$ are (in-)decomposable. PPTES are detected by indecomposable EWs \cite{Lewenstein:2000ab}.  
\end{enumerate}

The isomorphism between quantum channels and CJ operators shown in Eq. (\ref{eq:CJ3}) is useful to characterize properties of quantum states and channels. Suppose that, for a quantum channel $\Lambda$, the corresponding quantum state $\chi_{\Lambda} = (\i \otimes \Lambda) [ {P_{d_A}^{+}} ]$ is separable and also that it has the following separable decomposition, 
\bea
\chi_{\Lambda} = \sum_{i} p_i |e_i \rangle_A \langle e_i | \otimes  |f_i \rangle_B \langle f_i |. \nonumber
\eea
Note that if $[\i \otimes \Lambda] ({P_{d_A}^{+}})$ is separable, states $[\i \otimes \Lambda] (\rho)$ are also separable
for any state $\rho$ \cite{Horodecki:2003aa}. A quantum channel $\Lambda$ is called entanglement-breaking if for all states $\rho\in S(\H^{(A)} \otimes \H^{(B)})$ the resulting states $(\i \otimes \Lambda)[\rho]$ are separable \cite{Horodecki:2003aa}. From the isomorphism in Eqs. (\ref{eq:CJ2}) and (\ref{eq:CJ3}), one can find the corresponding map in the following form
\bea
\Lambda(\cdot) = \sum_i ~\tr[ (\cdot) ~p_i | e_{i}^{*} \rangle \langle e_{i}^{*} | ] ~ | f_i \rangle \langle f_i |. \label{eq:ebchannel}
\eea
This shows that, for a given state, the channel works as preparation of quantum state $\{|f_i\rangle \langle f_i | \}_i$ followed by measurement outcomes with POVMs $\{ p_i |e_{i}^{*}\rangle \langle e_{i}^{*}|  \}_i$. That is, entanglement-breaking channels can be implemented by a measure-and-prepare scheme. Conversely, a measure-and-prepare channel with rank-one POVMs is entanglement-breaking. 

It can happen that for a quantum channel $\Lambda$, its CJ operator is a bound entangled state \cite{Horodecki:1999aa}. The channel is called entanglement-binding. For instance, a channel $\Lambda$ is entanglement-binding if the CJ operator $\chi_{\Lambda}$ is PPTES. A quantum channel $\Lambda$ is called PPT-preserving if its CJ operator $\chi_{\Lambda} = (\i \otimes \Lambda) {P_{d_A}^{+}} $ is PPT, i.e., $\chi_{\Lambda}^T \geq 0$.

\begin{figure}
\begin{center}  
\includegraphics[width= 10cm]{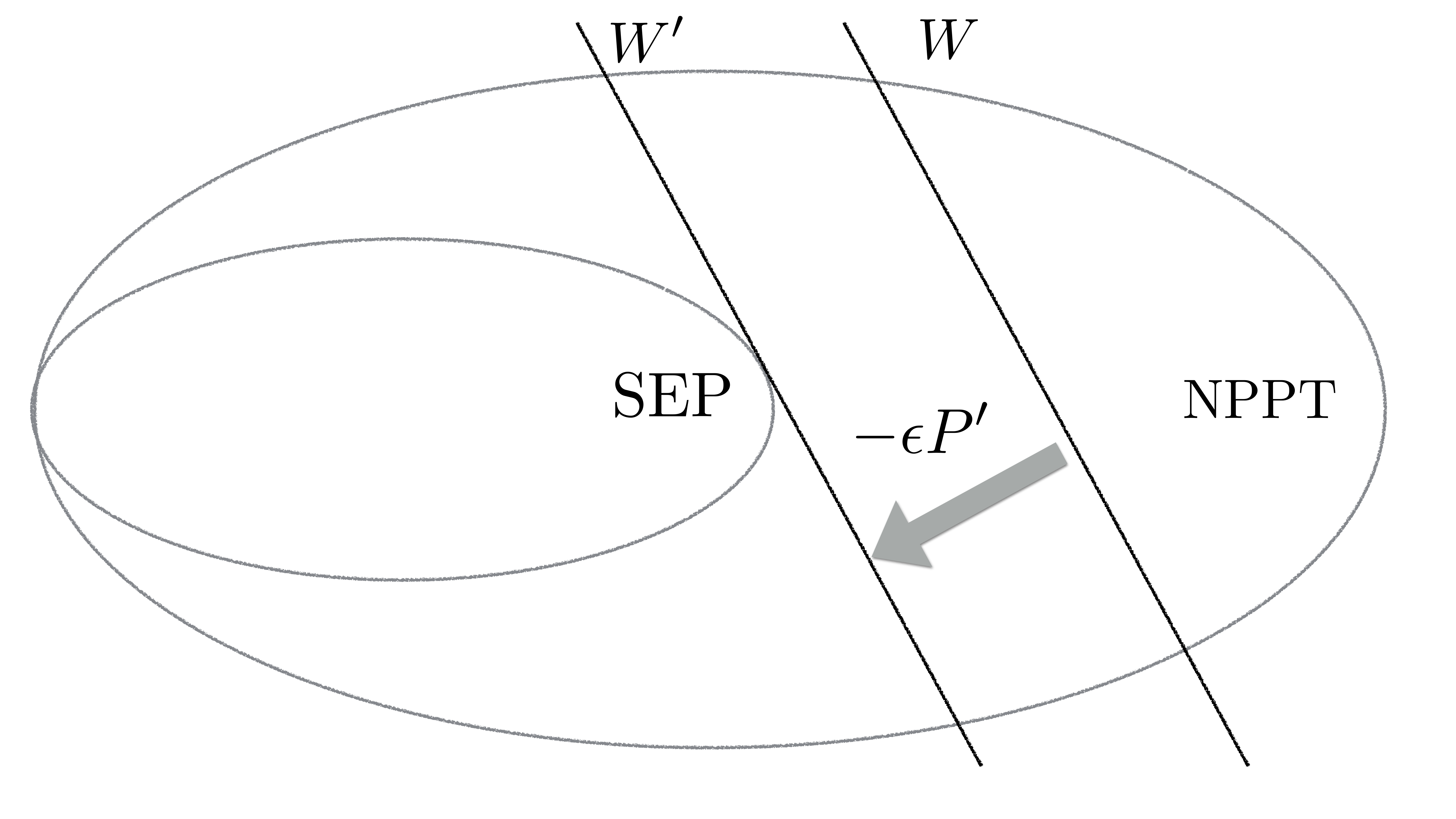}
\caption{ Optimization of EWs is depicted. A witness $W$ is optimized by subtracting a positive operator $P^{'}$, $ W \mapsto W^{'} = W - \epsilon P^{'}$. A resulting witnees $W^{'}$ is called (decomposable) optimal when there is no positive operator which can be subtracted further. After all, optimal EWs identify the border of separable states. } 
\label{fig:optimization}
\end{center}
\end{figure}

The isomorphism establishes the connection between positive but non-CP maps and EWs. In Eq. (\ref{eq:WL}), a positive but non-CP map can be either decomposable or indecomposable, by which one can find a canonical form of EWs as follows,
\bea
W = P + Q^{\Gamma}. \label{eq:witcan}
\eea
If $P\geq 0$, the EW can detect NPPT states and is called a decomposable EW. For some NPPT state $\rho$, we have $\tr[ \rho W] = \tr[ \rho P]  + \tr[  \rho^{\Gamma} Q]$, which is negative if $- \tr[ \rho^{\Gamma} Q] > \tr[  \rho P ] \geq 0$. 

Two EWs $W_i = P_i +Q^{\Gamma}$ for $i=1,2$ with $P_1 \geq P_2$ can be compared in terms of the detection capability \cite{Lewenstein:2000ab}. Denoted the set of detected states by $W$ as
\bea
D_W =\{ \rho \in S ( \H_A\otimes \H_B )~\| ~  \tr[ \rho W] <0\}, \nonumber 
\eea
one can find that $D_{W_1} \subset D_{W_2}$. That is, all states detected by $W_1$ are also detected by $W_2$. Then, $W_2$ is called finer than $W_1$. When an EW is finer than any other EWs, it is called optimal \cite{Lewenstein:2000ab}. When an EW is not optimal, optimization can be performed by subtracting positive operators: $W_1 \rightarrow  W' = W_1- \epsilon P'$ for some $P'\geq 0$ and $\epsilon \geq 0$. Then, EW $W$ is optimal if for any $P'\geq 0$ and arbitrary $\epsilon >0$, $W-\epsilon P'$ is no longer an EW. Collecting optimal EWs, all entangled states can be detected.

It is not immediate to find optimality of EWs. A useful method is the so-called spanning property, that is only sufficient condition for optimal EWs. Let $P_W$ denote the set of product states on which a witness $W$ vanishes,
\bea
P_W = \{  | e\rangle | f \rangle \in  \H^{(A)} \otimes \H^{(B)} ~ \| ~ \langle e |  \langle f| W'   | e\rangle| f\rangle =0  \}. \nonumber
\eea 
We call $W$ contains the spanning property if $P_W$ can span the whole space, i.e., $\mathrm{span} ~{P_W} =\H^{(A)} \otimes \H^{(B)}$. If an EW contains the spanning property, it is optimal. The converse, however, does not hold true in general.

In Eq. (\ref{eq:witcan}), EWs are called indecomposable if they are obtained from indecomposable positive maps. In this case, we have $P\ngeq 0$ in Eq. (\ref{eq:witcan}) and EWs can detect PPTES. Similarly, the optimization process works by subtracting decomposable operators, that are in the following form in general $D = P' + Q'^{\Gamma}$ with $P^{'}, Q^{'}\geq 0$. When there is no decomposable operator that can be subtracted, the resulting EWs are called non-decomposable-optimal. I.e., $W$ is non-decomposable optimal if $\forall \epsilon>0$ and for all decomposable operators $D$, it holds that $W - \epsilon D$ is no longer an EW \cite{Lewenstein:2000ab}.  

Then, from Eq. (\ref{eq:CJ1}) positive maps can be obtained from EWs, i.e. given $W$, the positive map can be derived as $\Lambda(\cdot) = d_A \tr[W ( (\cdot)^{T} \otimes\ido)  ]$. Positive maps are called (indecomposable) optimal if they are derived from (indecomposable) optimal EWs. EWs have been a useful tool to  investigate various structures and properties of entangled states. More properties and structures of EWs, such as optimality, extremality, atomicity, etc. have been explained in a recent  review \cite{Chruscinski:2014aa}. 

\subsection{Structural physical approximation and quantum channels } 

In this subsection, we discuss the relation between SPA and quantum channels, namely the conjecture addressed in Ref. \cite{SPAconjecture} that SPAed maps to optimal positive maps correspond to entanglement-breaking channels. While there have been a number of supporting examples \cite{SPAconjecture, Chruscinski:2009aa, Chruscinski:2010aa, Chruscinski:2011aa, Zwolak:2013aa, Zwolak:2014aa, Augusiak:2014aa}, counterexamples have been finally found, firstly in indecomposable cases \cite{Ha:2012aa, Stormer:2013aa} and then decomposable cases \cite{Chruscinski:2014ab}, see also numerical evidences \cite{Hansen:2015aa} and a recent review \cite{Shultz:2016aa}. In the following, we address the conjecture with an observation on no-go theorems in quantum theory and overview the progress. Positive maps that satisfy the conjecture are considered in detail and their practical applications are shown in the next section.

{\bf An observation on no-go theorems.} The conjecture has been motivated by an attempt to understanding disallowed dynamics in quantum theory. For instance, the no-cloning theorem states that unkonwn quantum states cannot be perfectly copied \cite{Wootters:1982aa, Scarani:2005aa}, which is closely related to the other, the impossibility of perfect state discrimination \cite{Helstrom, Holevo1974, Yuen:1975aa, Chefles:2000aa, Bergou:2004aa, Bergou2007, Barnett:2009aa, Bae:2015aa}. Note that the no-go theorems are the key elements in some of quantum information applications, e.g. quantum cryptographic protocols \cite{BB84, Bennett:1992aa}.

We now observe quantum operations that make approximations to disallowed dynamics, as follows. One can firstly consider optimal quantum cloning, for instance, the symmetric and universal $1\rightarrow 2$ cloning operation \cite{Buzek:1996aa}. It is clear that the $1\rightarrow 2$ quantum cloning is not an entangling-breaking channel. However, asymptotic quantum cloning where output clones tend to be sufficiently large, i.e., $N\rightarrow \infty$ quantum cloning, converges to entanglement-breaking \cite{Bae:2006aa, Chiribella:2006aa}. Application of asymptotic quantum cloning to bipartite quantum states gets rid of entangled states.

One can also consider another impossible operational task having different origin of the impossibility in quantum theory. Arbitrary manipulation of unknown quantum states is generally disallowed \cite{Hardy:2001aa}, among which the universal-NOT (UNOT) operation is an extreme case \cite{Buzek:1999aa, Gisin:1999aa, Bechmann-Pasquinucci:1999aa} and not possible either. The ideal UNOT operation works as
\bea
\forall |\psi\rangle \in S(\H_2),~~\Lambda_{ \mathrm{UNOT}}~: ~ |\psi \rangle ~ \mapsto~  |\psi^{\perp}\rangle,  \label{eq:UNOT} 
\eea
i.e., converting a state to its orthogonal complement. This is, however, an anti-unitary transformation that does not preserve a physical symmetry and consequently cannot be a legitimate quantum operation \cite{Wigner:1960aa}. In Ref. \cite{Buzek:1999aa}, a quantum channel that optimally approximates the UNOT operation has been shown. It turns out that the resulting quantum operation can be implemented by a measure-and-prepare protocol, that corresponds to an entanglement-breaking channel. We emphasize that the best approximate quantum operation turns out to be entanglement-breaking.

It is worth noting that the ideal UNOT operation is a positive but non-CP map since it can be rewritten as, from Eq. (\ref{eq:UNOT}),
\bea
\Lambda_{\mathrm{UNOT}} (~\cdot ~)= Y T[ ~ \cdot ~ ] Y \label{eq:TUNOT}
\eea
with the transpose map $T$ and Pauli matrix $Y$. This shows that the UNOT is equivalent to the transpose map up to a local unitary transformation. This immediately implies that for qubit states, SPA to the transpose is a measure-and-prepare scheme, that is, entanglement-breaking. This holds true in continuous-variable systems \cite{Buscemi:2003aa}. 

The result in Ref. \cite{Bechmann-Pasquinucci:1999aa} is particularly interesting. It shows that the optimal approximate UNOT coincides to what appears in the ancilla of the $1\rightarrow 2$ quantum cloning \cite{Buzek:1996aa}. To be precise, we note that the $1\rightarrow 2$ quantum cloning produces three qubits, in which two are approximate clones and the other is ancilla. The process happened in the ancila is named quantum anti-cloning, that coincides to an optimal approximate UNOT. In Ref. \cite{De-Martini:2002aa}, the approximate UNOT has been experimentally realized via the aforementioned anti-cloning process by implementing the $1\rightarrow 2$ quantum cloning. Technically, the approximating UNOT corresponds to the complementary channel of the  $1\rightarrow 2$ quantum cloning. Recall that the approximate UNOT operation is entanglement-breaking and can be equivalently implemented by a measure-and-prepare scheme \cite{Lim:2011aa}. \\

\begin{figure}
\begin{center}  
\includegraphics[width= 13cm]{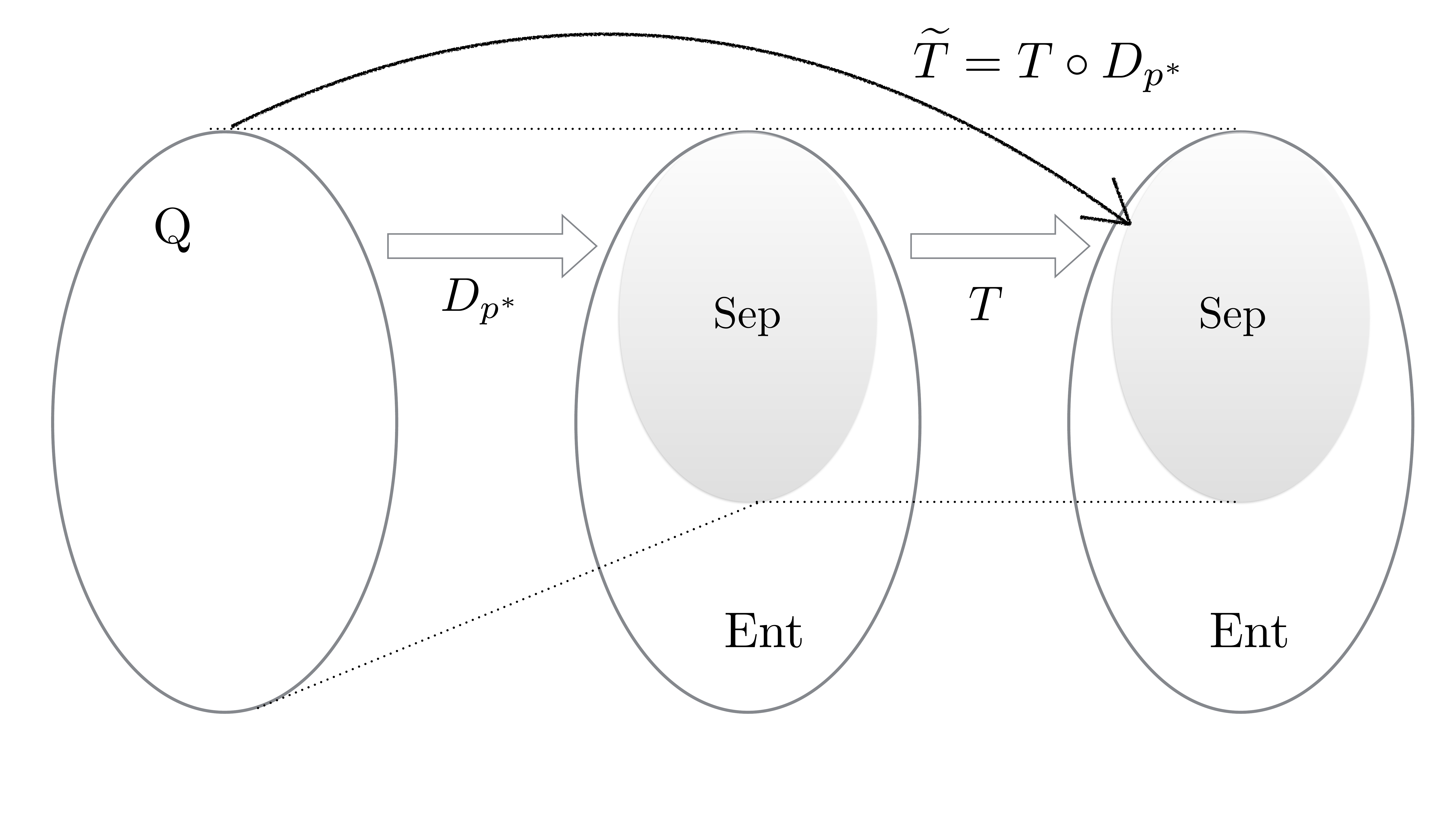}
\caption{ The SPAed transpose $\widetilde{T}$ is equal to the concatenation of the ranspose $T$ and the depolarization $D_{p^*}$, where one can show that the latter is entanglement-breaking. The interpretation is that the domain is reduced by the depolarization to separable states for which the transpose map is not only positive but also CP. } 
\label{fig:conjecture}
\end{center}
\end{figure}

{\bf The conjecture.} It has been observed that the approximate UNOT corresponds to an entanglement-breaking channel, by which all quantum states are mapped to separable states. From the fact that entanglement is closely connected to disallowed dynamics, UNOT, quantum cloning, the transpose map, etc., one may understand that approximations to the impossible tasks would be necessarily entanglement-breaking which gets rid of entangled  states. 

In fact, from Eq. (\ref{eq:spa}) the SPAed transpose can be written as $\widetilde{T} = T\circ D_{p^*}$ where $D_p (\rho)= (1-p^*)\rho + (p^* /d) \ido$ is an entanglement-breaking channel. The decomposition may elucidate the mathematical structure that the CP map $\widetilde{T}$ works by applying the transpose $T$ to separable states after the entanglement-breaking channel $D_{p^*}$, see Fig. \ref{fig:conjecture}. See also that the decomposition is referred to as $\mathrm{Positive}$-divisible map after properly including a parameter indicating time-evolution \cite{Chruscinski:2014ac, Breuer:2009aa}. In the range of the depolarization $D_{p^*}$, there are only separable states for which positive but non-CP maps have no reason to be regarded as non-physical ones. It can also be interpreted that, therefore, SPA to optimal positive maps should necessarily get rid of entangled states, hence, entanglement-breaking. The SPA conjecture generalizes the observation and has been addressed as follows \cite{SPAconjecture}. \\

{\it SPAs to optimal positive maps correspond to entanglement-breaking channels. Equivalently, SPAed optimal EWs are separable states.}\\

The conjecture can be tested in the following way. For a optimal positive map $\Lambda$, one has to firstly find the SPAed map $\widetilde{\Lambda}$ with $p^{*}$, see Eq. (\ref{eq:spa}). This can be obtained by finding minimal $p^*$ such that $(\id \otimes \Lambda)[P_{d}^{+}]\geq 0$ \cite{Horodecki:2003aa, SPAconjecture}. The non-trivial part is to check if the map $\widetilde{\Lambda}$ is entanglement-breaking or not. To do this, one has to determine if its CJ state $(\id \otimes \Lambda)[P_{d}^{+}]$ is separable, or not. If it is separable, it implies that the SPAed map is entanglement-breaking. Otherwise, the conjecture is disproven. 

Note that by admixing sufficient noise a non-CP map can always transformed to a CP map \cite{Augusiak:2011aa}. Let $p_s$ denote the noise parameter in Eq. (\ref{eq:spa}) instead of $p^*$ such that the resulting CP map is entanglement-breaking. It holds that $p_s \geq p^*$ in general, and the conjecture addresses the question if the equality holds for all optimal positive maps.

Interestingly, there have been numerous examples of positive maps that support the conjecture \cite{SPAconjecture, Chruscinski:2009aa, Chruscinski:2010aa, Chruscinski:2011aa, Zwolak:2013aa, Zwolak:2014aa, Augusiak:2014aa}. Recall that once a quantum channel is entanglement-breaking, there is a huge simplification in implementation that it can be performed by a measure-and-prepare scheme. Then, a quantum channel is designed as measurement of an input state and preparation of a quantum state according to measurement outcomes. In the next section \ref{section:spaexp}, structures of SPAed maps for positive maps satisfying the conjecture are shown and the detailed derivations are presented.  

Among the examples, a general property has been shown in Refs. \cite{Chruscinski:2010aa, Augusiak:2011aa}. Note that the class of isotropic states is given by $\rho_{\mathrm{iso}} = (1-p)P_{d}^{+} +  (p/d^2) \ido$, which is NPPT entangled when $p\in[0, d/(d+1))$ and otherwise separable. \\

{\bf Theorem.} If a positive map $\Lambda:\B(\H_d)\rightarrow \B(\H_d)$ detects all entangled isotropic states in the dimension, then its SPAed map $\widetilde{\Lambda}$ is entanglement-breaking. \\

This is useful when testing if a positive map satisfies the conjecture. In fact, it can be applied to the transpose, Reduction map, and some of Breuer-Hall maps \cite{SPAconjecture}. Note that Choi's map in Eq. (\ref{eq:choimap}) also satisfies the conjecture. We also remark that, interestingly, in the theorem above the optimality of positive maps is not necessary to fulfill the conjecture.

The conjecture has been extended to continuous-variable systems, and specially considered for Gaussian states \cite{Horodecki:2000aa}. In this case, the transpose map is particularly interesting since almost all entangled Gaussian states are NPPT, i.e., by the partial transpose significant fractions of entangled Gaussian states can be detected. The transpose map, denoted by $T_G$, works on the level of displacement operators \cite{PhysRevLett.84.2726, Duan:2000aa}. In Ref. \cite{Augusiak:2011aa}, it has been shown that the SPAed Gaussian channel $\widetilde{T}_{G}$ is entanglement-breaking, namely measurement in and preparation of coherent states, i.e.,
\bea
\widetilde{T}_{G} [\rho] = \frac{1}{\pi}  \int dx dy \langle x,y | \rho | x, y \rangle | x,-y \rangle \langle x, -y|  \nonumber
\eea
where $|x,y\rangle$ denotes a coherent state having displacements in $x$ and $y$.

Finally, the conjecture is disproven for both indecomposable \cite{Ha:2012aa, Stormer:2013aa} and decomposable cases \cite{Chruscinski:2014ab}, see also numerical evidences \cite{Hansen:2015aa}. We also refer to the excellent reviews on the mathematical structure of the conjecture and the counterexamples \cite{Shultz:2016aa} and on the detailed terms of optimality, extremality, atomicity, and facial structures of EWs \cite{Lewenstein:2000ab, Chruscinski:2014aa}. The counterexample for indecomposable cases is given in Ref. \cite{Ha:2012ab} that has been obtained by further generalizations of Choi maps in Ref. \cite{Cho:1992aa}. The map is given in $3$-dimensional Hilbert space: for non-negative real numbers $a$, $b$, and $c$ and $\theta\in [ -\pi, \pi]$ and $X = (x_{i,j})\in M_3$,
\bea
\Lambda[a,b,c,\theta] (X) = \left( \begin{array}{ccc} a x_{11} +b x_{22} + c x_{33} & -e^{i\theta} x_{12} & -e^{ -i \theta} x_{13} \\  -e^{-i\theta} x_{21} & cx_{11} + ax_{22} + bx_{33} & -e^{ i\theta} x_{23} \\ -e^{i\theta} x_{31} &  -e^{ - i\theta} x_{32} & b x_{11} +c x_{22} + a x_{33} \end{array}\right) \nonumber
\eea
With this example, the conjecture is tested and disproven in Ref. \cite{Ha:2012aa} where the conic properties of EWs have been completely analyzed. Namely, the SPAed map is not entanglement-breaking for $b>0$ and $\theta \in (-\pi/3,0)\cup (0,\pi/3)$. 


\setcounter{footnote}{0}
\section{Structural Physical Approximation for Practical Realization }
\markboth{\sc Structural Physical Approximation for Practical Realization}{}
\label{section:spaexp}

In the previous section, it is shown that SPA leads a number of positive maps into entanglement-breaking channels. We reiterate that for cases where SPAed maps are entanglement-breaking, the implementation works simply by measurement and preparation of quantum states, that are feasible with current technologies. In this section, we show how one can derive an implementation scheme for SPAed maps when the positive maps satisfy the conjecture. In particular, we take cases of the transpose map due to its usefulness for quantum information applications and the significance in fundamental aspect of quantum theory. The procedure can be applied to other positive maps  in general.  

We begin with a brief introduction to the transpose map. It was considered in the beginning of quantum theory when formulating legitimate quantum dynamics, and has been a standard example of anti-unitary transformations in quantum theory \cite{Wigner:1960aa}. Let $s$ denote  a symmetry transformation in quantum theory and $R(s)$ its representation. It satisfies the following, for any pair of states $|\psi\rangle$ and $|\phi\rangle$,
\bea
| \langle \phi | ~R^{\dagger}(\mathrm{s}) R(\mathrm{s})   (  | \psi \rangle )| = | \langle \phi |  \psi \rangle |
\eea
which corresponds to quantities that can be obtained in reality, for instance, expectations of observables and probabilities. Wigner has shown that the symmetry transformation must be given by either unitary or anti-unitary transformations: its unitary representation denoted by $R_U(\mathrm{s})$ and anti-unitary representation by $R_{A} (\mathrm{s})$ satisfies the followings, 
\bea
\mathrm{unitarity}&:&  R_U(\mathrm{s}) R_U(\mathrm{s})^{\dagger} =  R_U(\mathrm{s})^{\dagger} R_U(\mathrm{s}) = \ido, \nonumber \\
\mathrm{antiunitarity}&:& R_A(\mathrm{s}) R_A(\mathrm{s})^{\dagger} = R_A(\mathrm{s})^{\dagger}  R_A(\mathrm{s})=  - \ido. \nonumber
\eea
No experimental procedure can distinguish the two cases in the above. 

Extending to bipartite systems and symmetry transformations $\mathrm{s}$ and $\mathrm{s}^{'}$, it holds that both $R_U(\mathrm{s}) \otimes R_{U}(\mathrm{s}^{'})$ and $R_A(\mathrm{s}) \otimes R_{A}(\mathrm{s}^{'})$ are unitary. However, tensor product of unitary and anti-unitary transformations $R_U(\mathrm{s}) \otimes R_A(\mathrm{s}^{'})$ is neither unitary nor anti-unitary, hence, no longer forms a symmetry transformation, that may lead to some non-physical features. Note that an anti-unitary transformation can be decomposed into successive applications of unitary transformation and the transpose, i.e. an anti-unitary transformation $U_{\mathcal{A}}$ has a decomposition $U_{\mathcal{A}} = T \circ U$ for some unitary transformation $U$ \cite{Wigner:1960aa}. Then, one can find non-physical maps, tensor product of unitary and anti-unitary transformations, have the following decomposition, 
\bea
R_U(\mathrm{s}) \otimes R_A(\mathrm{s}^{'}) = (\i \otimes T)\circ (R_U(\mathrm{s}) \otimes R_U(\mathrm{s}^{'})) \label{eq:UAUtoPT}
\eea
for some unitary representation of $\mathrm{s}^{'}$. 

Since the term $R_U(\mathrm{s}) \otimes R_U(\mathrm{s}^{'})$ of the right-hand-side in Eq. (\ref{eq:UAUtoPT}) is unitary, it is concluded that the partial transpose $\i \otimes T$ is the origin that the tensor product of unitary and anti-unitary transformations gives rise to a non-physical phenomenon. This immediately means that the transpose {\it per se} is not physical: no physical process corresponds to the transpose. That is to say, there does not exist a positive and CP map that describes the transpose. 

In what follows, we present implementation schemes and proposals for the SPA to transpose map for qubit, qutrit, and qudit states. We also show an LOCC protocol that realizes the SPA to the partial transpose.

\subsection{Transpose }


{\bf Qubit states.} We begin with the SPA to the transpose for qubit states and provide the derivation in detail. For qubit states, the transpose can be defined without dependence on the choice of basis. For convenience, let us take the transpose with respect to computational basis as follows,
\bea
T~: ~ \rho = \left[   \begin{array}{ccc} \rho_{11} & \rho_{12}  \nonumber \\  \rho_{21}  & \rho_{22}  \end{array} \right] ~\mapsto ~\rho^T =   \left[   \begin{array}{ccc} \rho_{11} & \rho_{21} \nonumber \\    \rho_{12}  & \rho_{22}  \end{array} \right]. \nonumber 
\eea
Note that the transpose itself is a linear and positive map. The SPA to the transpose is then obtained from the construction in Eq. (\ref{eq:spa}), 
\bea
\widetilde{T} =  \frac{1}{3} T + \frac{1}{3}D, \label{eq:T2SPA}
\eea
for which it holds that $\widetilde{T}$ is completely positive, i.e. $ (\i \otimes  \widetilde{T} ) (|\phi_{2}^{+} \rangle \langle \phi_{2}^{+} | ) \geq 0$. Note that it is found $p^{*} = 2/3$.  

To realize the SPAed transpose $\widetilde{T}$ in a prepare-and-measure scheme, one can exploit the isomorphism between channels and states. Given the map $\widetilde{T}$, the CJ operator can be found as,
\bea
\chi_{\widetilde{T}} = \frac{1}{3} (|\phi^{+}\rangle \langle \phi^{+} | + |\phi^{-}\rangle \langle \phi^{-} | + |\psi^{+}\rangle \langle \psi^{+} |) \label{eq:jamiT2}
\eea
where $|\phi^{\pm}\rangle = (|00\rangle \pm |11\rangle )/\sqrt{2}$ and $|\psi^{\pm} \rangle = (|01\rangle \pm | 10\rangle) /\sqrt{2}$. Note that the state in the above is separable \cite{Werner:1989aa}.Then, from the isomorphism between states and channels, one can find the channel performing the SPA to the transpose as follows, see Eq. (\ref{eq:CJ3})
\bea
\widetilde{T} [\rho] & = & 4 \tr_{12} [ \rho_1 \otimes (\chi_{\widetilde{T}})_{23}  | \phi^{+} \rangle_{12}\langle \phi^{+} | ] \label{eq:choit}\\
& = & \frac{1}{3} (\rho + X \rho X + Z \rho Z ) \label{eq:tch}
\eea
where $X$ and $Z$ denote Pauli matrices. This shows that an optimal approximation to the transpose can be achieved by the a random unitary channel shown in Eq. (\ref{eq:tch}). Since the CJ operator in Eq. (\ref{eq:jamiT2}) is separable, it is immediately shown that the SPAed transpose must be entanglement-breaking. However, the channel in Eq. (\ref{eq:tch}) is not yet in a form of a measure-and-prepare scheme. For the purpose, one has to find a separable decomposition of the state in Eq. (\ref{eq:choit}), that is in general highly non-trivial.

For two-qubit states, the results in Ref. \cite{Wootters:1998aa} can be used to find a separable decomposition. Note that in Ref. \cite{Wootters:1998aa}, the infimum of entanglement of formation for two-qubit states is obtained by finding a separable decomposition that optimally approximates the infimum. The technique can be applied as follows. Suppose that a separable decomposition for the CJ operator is given by the following form,
\bea
\chi_{\widetilde{T}} = \frac{1}{3} (|\phi^{+}\rangle \langle \phi^{+} | + |\phi^{-}\rangle \langle \phi^{-} | + |\psi^{+}\rangle \langle \psi^{+} |) = \sum_{i=1}^4 |z_i\rangle \langle z_i | \label{eq:sepdecomWerner}
\eea
with states $\{ |z_i\rangle \}_{i=1}^4$
\bea
|z_1\rangle & = & \frac{1}{2} ( e^{i\theta_1 } |x_1\rangle +  e^{i\theta_2 } |x_2\rangle  +  e^{i \theta_3 } |x_3 \rangle  + e^{i \theta_4 } |x_4\rangle ) \nonumber \\
|z_2\rangle & = & \frac{1}{2} ( e^{i\theta_1 } |x_1\rangle +  e^{i\theta_2 } |x_2\rangle  -  e^{i \theta_3 } |x_3 \rangle  - e^{i \theta_4 } |x_4\rangle ) \nonumber \\
|z_3\rangle & = & \frac{1}{2} ( e^{i\theta_1 } |x_1\rangle -  e^{i\theta_2 } |x_2\rangle  +  e^{i \theta_3 } |x_3 \rangle  - e^{i \theta_4 } |x_4\rangle ) \nonumber \\
|z_4\rangle & = & \frac{1}{2} ( e^{i\theta_1 } |x_1\rangle -  e^{i\theta_2 } |x_2\rangle  -  e^{i \theta_3 } |x_3 \rangle  + e^{i \theta_4 } |x_4\rangle ) \nonumber 
\eea
where $\{ |x_i\rangle\}_{i=1}^4$ are the so-called magic basis
\bea
| x_1 \rangle = i \sqrt{ \frac{1}{3} } | \psi^{-}\rangle, &~~~~~& | x_2 \rangle = \sqrt{  \frac{1}{2} } | \psi^{+ }\rangle \nonumber \\
| x_3 \rangle =  \sqrt{  \frac{1}{2} } | \phi^{-}\rangle, &~~~~~& | x_4 \rangle = i \sqrt{ \frac{1}{2}} | \phi^{+ }\rangle. \nonumber 
\eea
Note that the phase parameters $\theta_i$ for $i=1,2,3,4$ should satisfy the following condition
\bea
e^{-2 i \theta_2} + e^{ -2 i \theta_3 } + e^{ -2 i \theta_4  }  =0. \label{eq:theta}
\eea
The phase parameters are not uniquely determined, which also means that a separable decomposition is not unique. For convenience, we take as an instance $\theta_2=0$, $\theta_3 = \pi/3$ and $\theta_4 = -\pi  / 3$ that satisfy the relation in the above, and then obtain a separable decomposition $|z_i\rangle = | v_i\rangle  | v_i\rangle $ for $i=1,2,3,4$ in Eq. (\ref{eq:sepdecomWerner}) given with the following states,
\begin{eqnarray}
 \left| {v_1 } \right\rangle  &=& \sqrt {\frac{{3 - \sqrt 3 }}{6}} \left( {\left| 0 \right\rangle  + \frac{{ie^{i\pi 2/3} }}{{i + e^{ - i\pi 2/3} }}\left| 1 \right\rangle } \right) \nonumber \\
 \left| {v_2 } \right\rangle  &=& \sqrt {\frac{{3 + \sqrt 3 }}{6}} \left( {\left| 0 \right\rangle  - \frac{{ie^{i\pi 2/3} }}{{i + e^{ - i\pi 2/3} }}\left| 1 \right\rangle } \right) \nonumber\\
 \left| {v_3 } \right\rangle  &=& \sqrt {\frac{{3 + \sqrt 3 }}{6}} \left( {\left| 0 \right\rangle  + \frac{{ie^{i\pi 2/3} }}{{i - e^{ - i\pi 2/3} }}\left| 1 \right\rangle } \right) \nonumber\\
 \left| {v_4 } \right\rangle  &=& \sqrt {\frac{{3 - \sqrt 3 }}{6}} \left( {\left| 0 \right\rangle  - \frac{{ie^{i\pi 2/3} }}{{i + e^{ - i\pi 2/3} }}\left| 1 \right\rangle } \right). \label{eq:tetrastate}
\end{eqnarray}
Hence, a separable decomposition of the CJ operator in Eq. (\ref{eq:sepdecomWerner}) is obtained.

\begin{figure}
\begin{center}  
\includegraphics[width= 8cm]{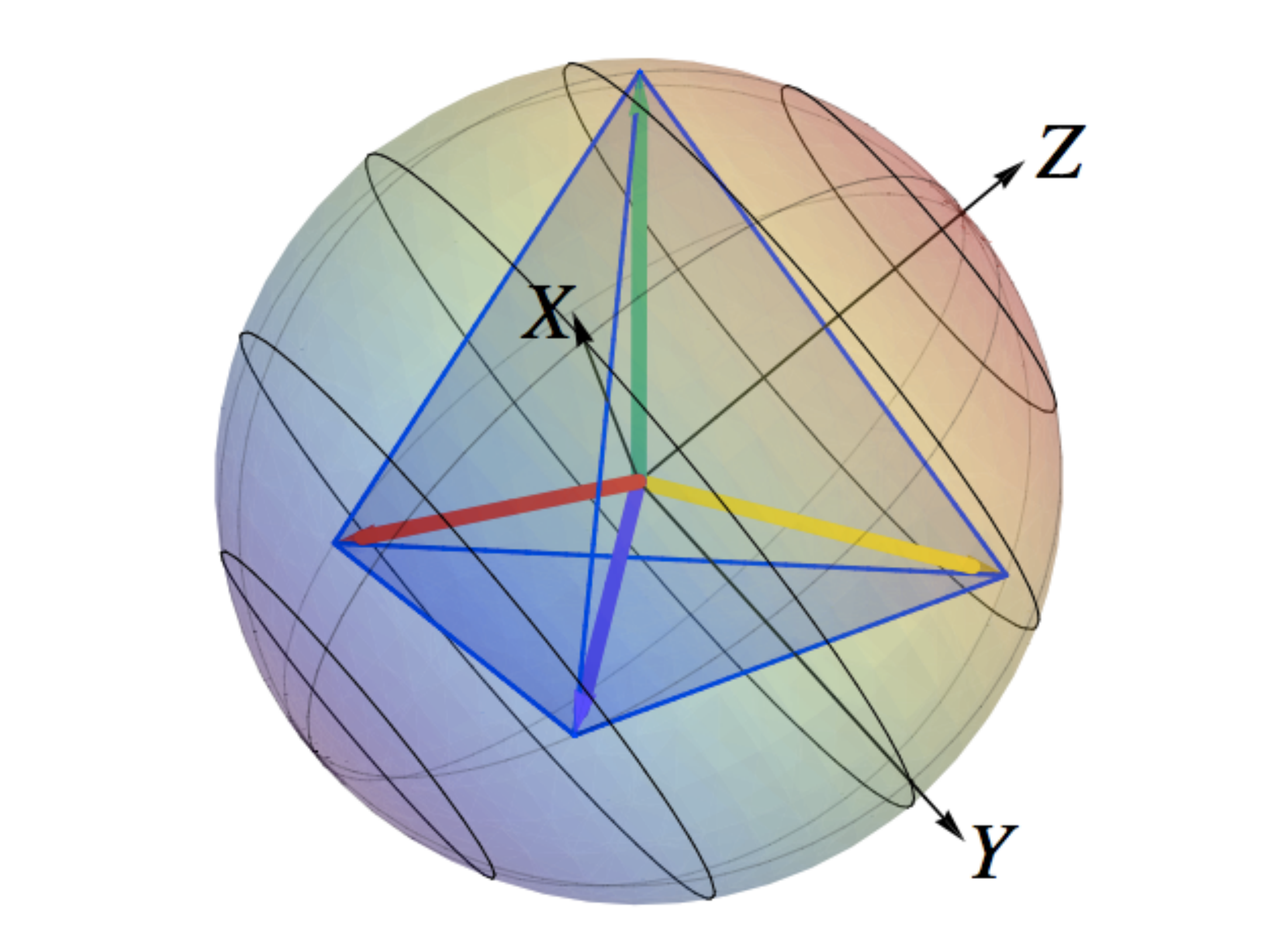}
\caption{ The four states $\{ |v_k\rangle  \}_{k=1}^4 $ in Eq. (\ref{eq:tetrastate}) form a tetrahedron in the Bloch sphere. The SPAed transpose in Eq. (\ref{eq:T2SPA}) can be realized in measurement in the conjugation $\{ |v_{k}^* \rangle \langle v_{k}^* |  /2 \}_{k=1}^4 $ and preparation in the tetrahedron states $\{ |v_{k}\rangle  \}_{k=1}^4 $. } 
\label{fig:tetra}
\end{center}
\end{figure}

Before moving to devising a measure-and-prepare scheme, we remark that  the four states $\{ |v_i\rangle \}_{i=1}^4 $ form a tetrahedron in the Bloch sphere, see Fig. \ref{fig:tetra}. Note also that by choosing different combinations of $\{ \theta_i\}_{i=1}^4$ in Eq. (\ref{eq:theta}), the tetrahedron rotates within the Bloch sphere while the CJ operator in Eq. (\ref{eq:jamiT2}) is kept unchanged. In high-dimensions, this property is generalized to quantum two-design \cite{KalevBae}. 

A measure-and-prepare scheme for the SPAed transpose can be obtained from Eq. (\ref{eq:choit}) and is given as follows, 
\bea
\widetilde{T} [\rho] = \sum_{i=1}^4 \tr[\frac{1}{2} | v_{i}^{*}   \rangle \langle v_{i}^{*}  |~ \rho ]  ~| v_{i} \rangle \langle v_{i}  | =   \sum_{i=1}^4 K_i \rho   K_{i}^{\dagger} \label{eq:spatm}
\eea
where $\{ K_i = \sqrt{2}^{-1} |  v_i \rangle \langle v_{i}^{*}   |  \}_{i=1}^4 $ are Kraus operators. This shows that the approximate transpose can be implemented by measurement $\{ M_i =  \frac{1}{2} |v_{i}^{*}  \rangle \langle v_{i}^{*}  |   \}_{i=1}^4 $ and preparation of states $\{ |v_i\rangle \langle v_i |  \}_{i=1}^4$: measurement outcome in basis $|v_{i}^{*} \rangle$ prepares its transposed one $|v_{i} \rangle$ for i=1,2,3,4, respectively. 

\begin{figure}
\begin{center}  
\includegraphics[width= 14cm]{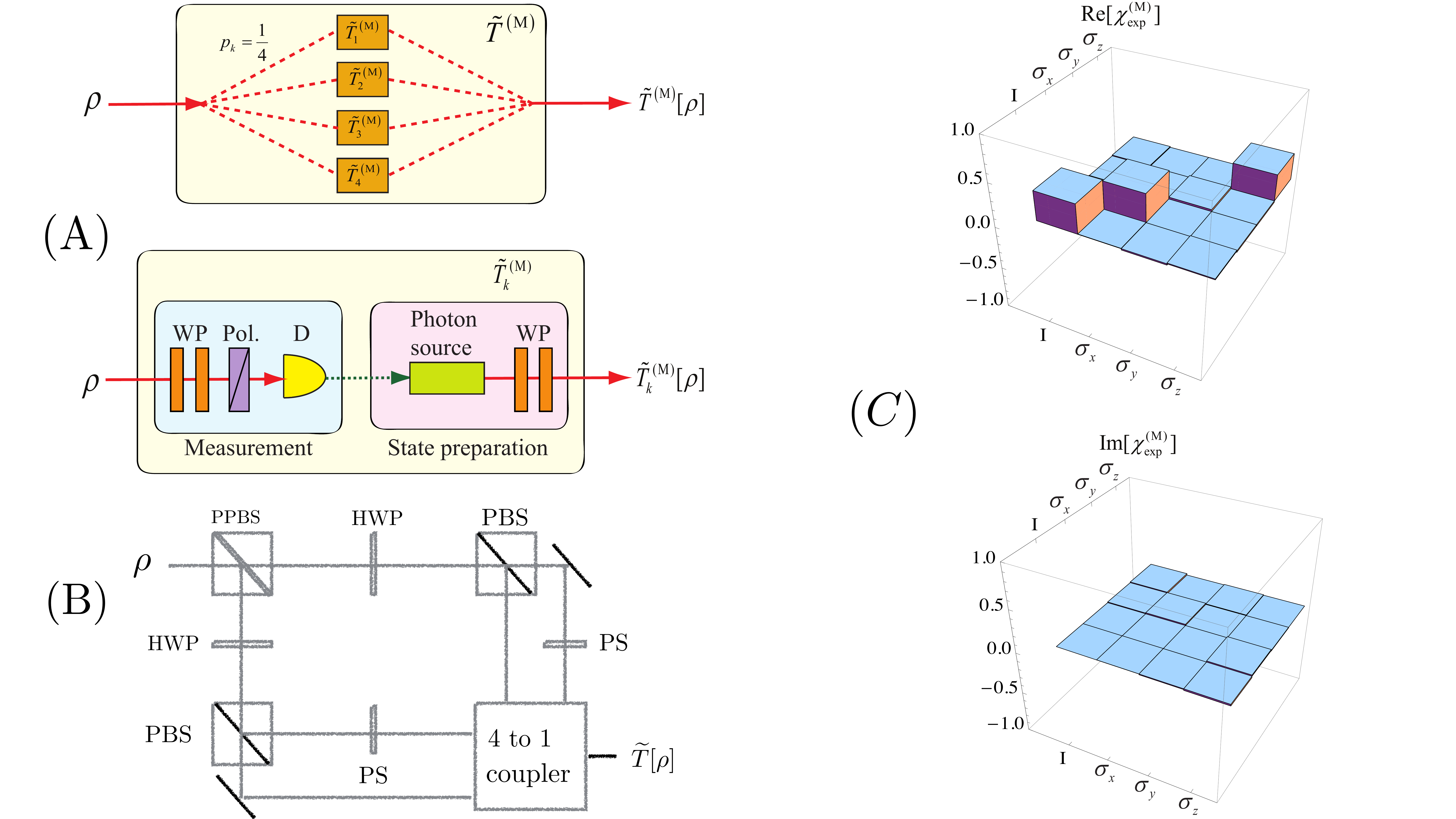}
\caption{ Implementation of the SPAed transpose for polarization qubit states is shown. (A) A proof-of-principle demonstration is performed for the SPAed transpose. WP, Pol. and D denote wave plate, polarizer, and detector, respectively. They are arranged to prepare basis $\{|v_{i}^* \rangle\}$ and $\{|v_{i} \rangle\}$ in Eq. (\ref{eq:tetrastate}), see more details in Ref. \cite{Lim:2011aa}. (B) Whereas (A) is a proof-of-principle demonstration, a trace-preserving scheme polarization qubit states can also be devised, where PPBS denotes partially-polarized beam splitter, HWP for half-wave plate, and PS for phase shifter. The scheme implements $\{|v_{i}^* \rangle\}$ and $\{|v_{i} \rangle\}$ in Eq. (\ref{eq:tetrastate}) via the construction of the states with Heisenberg-Weyl group, see Ref. \cite{KalevBae} for the details. (C) The quantum process tomography for the experiment that performs scheme (A) is obtained \cite{Lim:2011aa}, see Eq. (\ref{eq:jamiT2}) for its CJ operator. } 
\label{fig:qubitTsetup}
\end{center}
\end{figure}

The prepare-and-measure scheme for the SPAed transpose, obtained in Eq. (\ref{eq:spatm}), has been performed with polarization qubit states \cite{Lim:2011aa}, see also Fig. \ref{fig:qubitTsetup}, as a proof-of-principle demonstration. In polarization qubit states, the computational basis are identified by horizontal and vertical polarizations of single photons, $|0\rangle = |H\rangle$ and $|1\rangle  = |V \rangle$. In Fig. \ref{fig:qubitTsetup} (C), the process tomography is shown, from which one can also conversely find the random-unitary channel in Eq. (\ref{eq:tch}). 

In fact, the SPAed transpose is equivalent to the UNOT operation \cite{Buzek:1999aa} up to local unitary transformation $Y$, as it is shown in Eq. (\ref{eq:TUNOT}). The approximate UNOT operation has been realized in Ref. \cite{De-Martini:2002aa} as what happens in the ancillas of $1\rightarrow 2$ quantum cloning, named the anti-cloning process \cite{Bechmann-Pasquinucci:1999aa}. The experimental implementation stimulated emission in parametric down conversion that is a natural way of realizing $1\rightarrow 2$ quantum cloning \cite{De-Martini:1998aa, Simon:2000aa, De-Martini:2000aa, Lamas-Linares:2002aa}.

\begin{figure}
\begin{center}  
\includegraphics[width= 11.5cm]{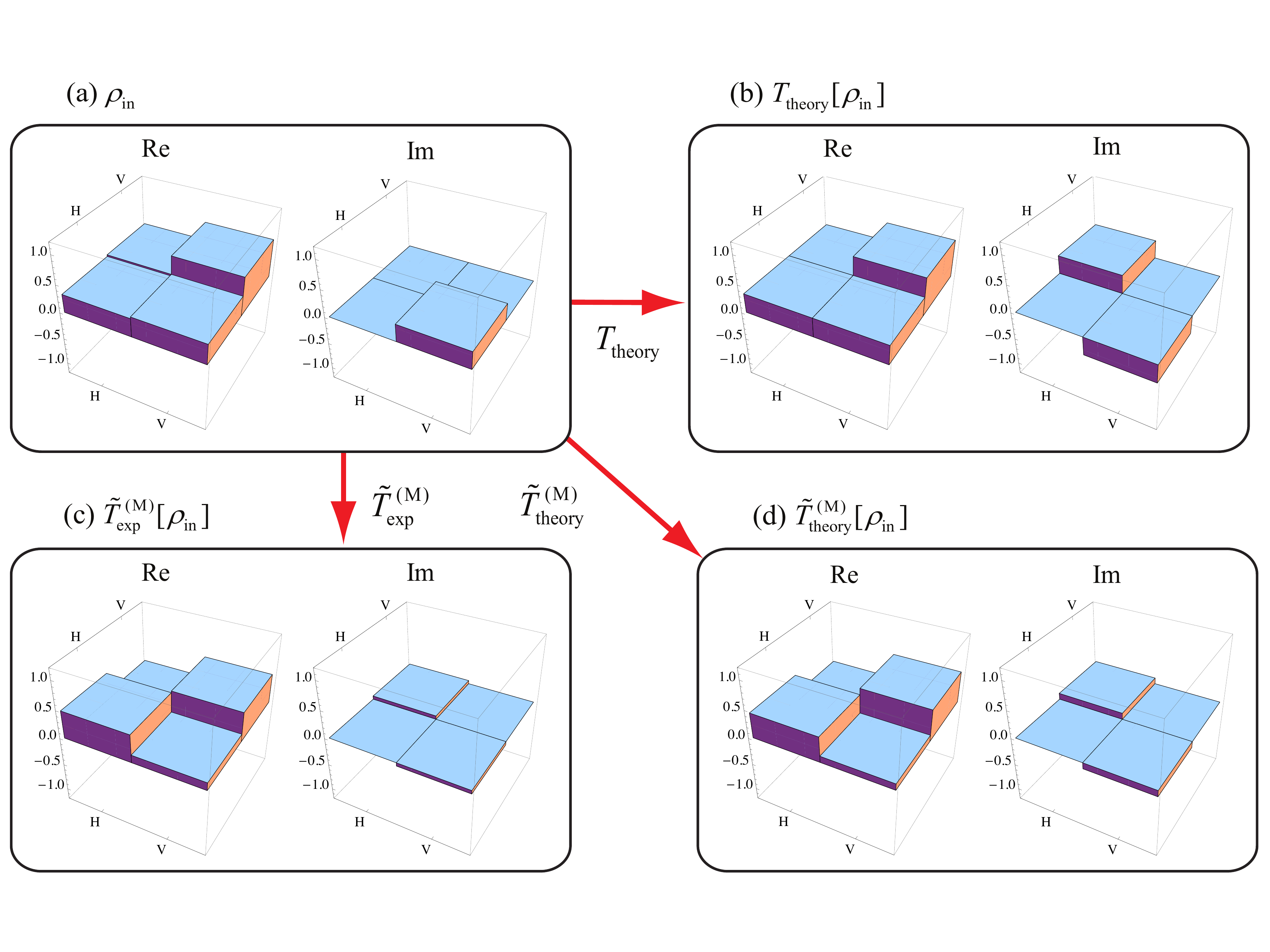}
\caption{For a state randomly prepared in (a), the SPAed transpose is applied with the scheme in Fig. \ref{fig:qubitTsetup} (A) \cite{Lim:2011aa}. (b) shows the ideal transpose and (c) is obtained by the SPAed transpose in experiment. The latter can be compared with (d), the ideal case of the SPAed transpose, and the gate fidelity is about $99\%$. }
\label{fig:qubitTexp}
\end{center}
\end{figure}

In Fig. \ref{fig:qubitTsetup} (A), For a given state $\rho$ the operation is performed by applying the channels $\widetilde{T}_{i}^{(M)}$ for $i=1,\cdots,4$ with probabilities $1/4$, respectively, where each $\widetilde{T}_{i}^{(M)}$ implements measurement in basis $| v_{i}^* \rangle\langle v_{i}^* |$ and preparation on the state $| v_{i} \rangle\langle v_{i} |$. Then, the resulting state on average is found in $\widetilde{T} [\rho]$. A single-shot scheme is shown in Fig. \ref{fig:qubitTsetup} (B) \cite{KalevBae}, 

As an instance, the following state is randomly generated and then identified by quantum state tomography,
\bea
\rho_{in} = \left( \begin{array}{ccc} 0.322 & 0.352- 0.307i  \nonumber \\  0.352 + 0.307i  & 0.678 \end{array} \right) ~~ \label{eq:exstate}
\eea
which is close to a pure state, e.g., one can check that $\tr[\rho_{in}^2] \approx 1$. In Fig. \ref{fig:qubitTexp}, the three cases are compared i) the ideal operation $T[\rho_{\mathrm{in}}]$ in $(b)$, ii) the ideal and physical one $\widetilde{T}_{\mathrm{theory}}[\rho_{\mathrm{in}}]$ in $(d)$, and iii) the experimental result of the realization $\widetilde{T}_{\mathrm{exp}}[\rho_{in}]$ in $(c)$. In Ref. \cite{Lim:2011aa}, it is reported that the obtained fidelity between $\widetilde{T}_{\mathrm{theory}}[\rho_{\mathrm{in}}]$ and $\widetilde{T}_{\mathrm{exp}}[\rho_{in}]$ is higher than $0.99$. All these are obtained by the scheme in Fig. \ref{fig:qubitTsetup} (A), for which the process tomography is also obtained in Fig. \ref{fig:qubitTsetup} (C).\\

{\bf Qudit states.} In the SPAed transpose for qubit states, we have observed that the measure-and-prepare scheme is performed with a set of tetrahedron states, see Fig. \ref{fig:tetra}. This follows from Eq. (\ref{eq:jamiT2}) that a separable decomposition of the CJ operator corresponds to the projection onto symmetric subspace of $\B(\H_2 \otimes \H_2)$. This in fact makes it feasible to find a separable decomposition of the separable CJ operator and leads to the prepare-and-measure scheme. 

The aforementioned properties of the SPAed transpose for qubit states can be generalized to high dimensional cases by relating the symmetric projection to quantum two-design \cite{KalevBae}. Let first us introduce projections onto symmetric $S_d$ and anti-symmetric subspaces $A_d$ as follows \cite{Dur:2000aa}
\bea
S_d = \frac{1}{2} (\ido + \Pi_d)~ \mathrm{and}~A_d  = \frac{1}{2} (\ido - \Pi_d) \nonumber
\eea
where $\Pi_d$ denotes the permutation operator in $\B(\H_d \otimes \H_d)$, i.e. $\Pi_d = \sum_{i,j =1}^d | i j \rangle \langle j i |$. Then, for the maximally entangled state $| \phi_{d}^+ \rangle$,  one can find the following relation $d  | \phi_{d}^+ \rangle \langle \phi_{d}^+ |^{\Gamma_B} =  (S_d - A_d)$. 

To obtain the SPAed transpose for a $d$-dimensional system, see Eq. (\ref{eq:spa}), we compute the value $p^*$ such that the map $\widetilde{T}$ is completely positive. This leads to the following:  
\bea
 && (\i \otimes \widetilde{T}) [|\phi_{d}^+ \rangle \langle \phi_{d}^+  |] \geq 0 \nonumber \\
& \Longleftrightarrow &   (p+d(1-p)) S_d + (p-d(1-p))  A_d\geq 0. \label{eq:SPATqudit}
\eea
Note that the anti-symmetric projection can be described as $A_d = \sum_{i>j} |\psi_{ij}^{-}\rangle \langle \psi_{ij}^{-} |$ with $|\psi_{ij}^{-} \rangle = ( | ij\rangle -| ji \rangle   )/ \sqrt{2}$. From the condition in the above, it is straightforward that the minimal $p$ satisfying in Eq. (\ref{eq:SPATqudit}) is given as,
\bea
p\geq p^* = \frac{d}{d+1}. \nonumber
\eea
From Eq. (\ref{eq:spa}), the SPAed transpose for $d$-dimensional quantum systems and its CJ operator are obtained as 
\bea
\widetilde{T} & = &  \frac{1}{d+1} T + \frac{d}{d+1} D, \nonumber
\eea
and its CJ operator is straightforwardly obtained as 
\bea
\chi_{\widetilde{T}} & =&  (\i \otimes \widetilde{T}) [|\phi_{d} \rangle \langle \phi_{d}  |] = \frac{2}{d(d+1)} S_d. ~~~\label{eq:td}
\eea
This shows that the CJ operator corresponds to the projection onto the symmetric subspace. 

In fact, a projection onto the symmetric subspace defines quantum two-design. Let us briefly summarize the terminology as follows. A set of states $\{ | x_k\rangle \in S(\H)\}_{k=1}^N$ is called spherical $t$-design if it holds that 
\bea
\mathrm{Sym}(\H^{\otimes t}) = \frac{1}{N} \sum_{k=1}^N  | x_k \rangle \langle x_k |^{\otimes t} \nonumber 
\eea
where $\mathrm{Sym(\H^{\otimes t})}$ denotes a symmetric subspace on space $\H^{\otimes t}$. The CJ operator in Eq. (\ref{eq:td}) is the case of spherical two-design, $t=2$.

It has been found that mutually unbiased bases (MUBs) \cite{Ivonovic:1981aa, Wootters:1989aa} and symmetric, informationally complete (SIC) states \cite{Renes:2004aa, ZAUNER:2011aa} are spherical two-design. That is, the CJ operator in Eq. (\ref{eq:td}) has separable decompositions with MUBs and SIC states. In a $d$-dimensional Hilbert space, if there exist MUBs there are $d+1$ MUBs, and if SIC states exist there are $d^2$ SIC states. Note that the existence of these states is a longstanding open problem in quantum information theory \cite{ Englert:2001aa, ZAUNER:2011aa}. 

When MUBs or SIC states exist, let $\{ |b_{j}^{k} \rangle \}_{j=1}^{d}$ for $k=1,\cdots,d+1$ denote MUB and $\{|s_i \rangle \}_{i=1}^{d^2} $ SIC states. For dimensions where MUBs exist, it holds that 
\bea
\forall~k\neq k^{'}=1,\cdots,d+1,~~ |  \langle b_{i}^k | y_{j}^{k^{'}} \rangle |^2 = \frac{1}{d}, \nonumber
\eea
for all $i,j$. For instance, when $d=2$, MUBs are eigenstates of Pauli matrices,
\bea
\mathrm{z-basis}~~~~~~  |0\rangle,&&  |1\rangle, \nonumber \\
\mathrm{x-basis}~~~~~~  \frac{1}{\sqrt{2}}(|0\rangle + |1\rangle), && \frac{1}{\sqrt{2}}(|0\rangle - |1\rangle), \nonumber \\
\mathrm{y-basis}~~~~~~  \frac{1}{\sqrt{2}}(|0\rangle + i|1\rangle), && \frac{1}{\sqrt{2}}(|0\rangle -i |1\rangle)  \label{eq:MUB2}
\eea
For dimensions where SIC states exist, SIC states satisfy the following relation,
\bea
\forall i\neq j, ~~ | \langle s_i | s_j \rangle |^2 = \frac{1}{d+1}. \nonumber
\eea
For instance, when $d=2$, any set of four states forming a tetrahedron in the Bloch sphere is an instance of SIC states, see also Eq. (\ref{eq:tetrastate}).

Then, in dimensions where MUBs or SIC states exist, the CJ operator in Eq. (\ref{eq:td}) has separable decompositions 
\bea
\mathrm{MUB ~construction } &: & \frac{1}{d(d+1)} \sum_{k=1}^{d+1} \sum_{j=1}^{d }  | b_{j}^{k}\rangle \langle b_{j}^{k} | \otimes | b_{j}^{k}\rangle \langle b_{j}^{k} |  \nonumber \\
\mathrm{SIC ~construction }& : & \frac{1}{d^2} \sum_{j=1}^{d^2} | s_j \rangle \langle s_j | \otimes |s_j\rangle \langle s_j |. \nonumber
\eea
When separable decompositions are given as above, the SPAed transpose can be implemented as measurement and preparation in MUBs or SIC states:
\bea
&&\mathrm{MUB~construction: } \nonumber \\
&&\widetilde{T} [\rho] = \frac{1}{d(d+1)} \sum_{i=1 }^{d} \sum_{k =1}^{d+1} \tr [  \frac{1}{d(d+1)}    |b_{j}^{k*}\rangle \langle b_{j}^{k*} | \rho ]   ~ |b_{j}^k \rangle \langle b_{j}^k |  \label{eq:MUBT} \\
&& \mathrm{SIC~construction: } \nonumber \\
~~~~~~&&\widetilde{T} [\rho] = \frac{1}{d } \sum_{i=1 }^{d^2} \tr [  \frac{1}{d } | s_{j}^{*}\rangle \langle s_{j}^{*} |  \rho ]    ~  |s_j\rangle \langle s_j |. \label{eq:sicT}
\eea
For instance, the SPAed transpose for qubit states in Eq. (\ref{eq:spatm}) with SIC states can equivalently implemented with MUBs as follows,
\bea
\widetilde{T} [\rho] = \frac{1}{6} \sum_{k = x,y,z} \tr[ | 0^{* } \rangle_k\langle 0^{*} | \rho]   | 0\rangle_k\langle 0|    + \tr[ |  1^{* } \rangle_k\langle 1^{*} |  \rho]  | 1 \rangle_k\langle 1 |  \nonumber
\eea
where $|i\rangle_k$ for $i=0,1$ denotes the computational basis in $k$-axis, see Eq. (\ref{eq:MUB2}) for $k=x,y,z$. That is, measurement in computational basis in all directions and preparation of their conjugate states construct the SPAed transpose. In principle, the implementation schemes presented in Eqs. (\ref{eq:MUBT}) and (\ref{eq:sicT}) can be generally applied to $d$-dimensional quantum systems where MUBs or SIC states exist.

\begin{figure}
\begin{center}  
\includegraphics[width= 12cm]{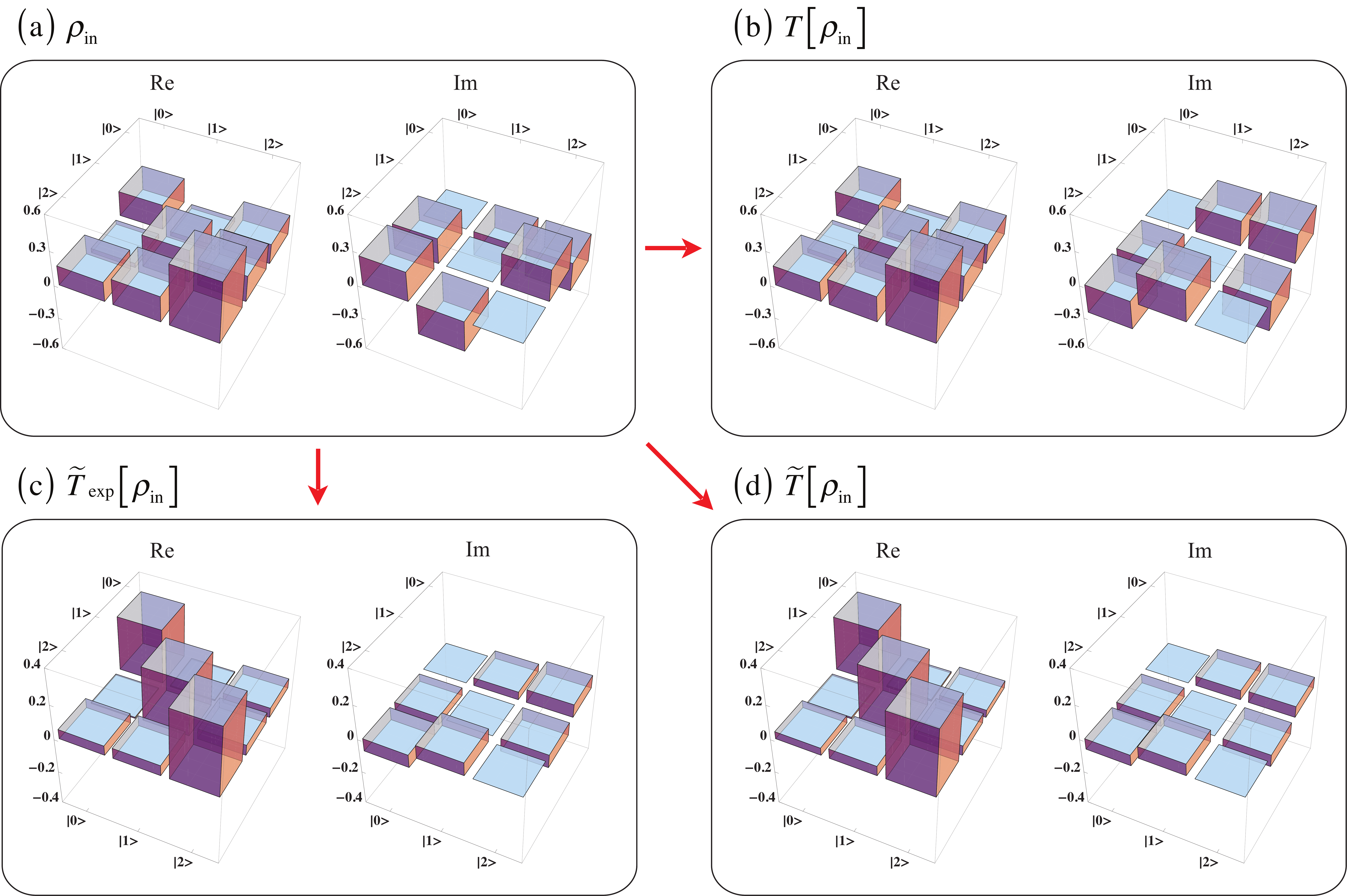}
\caption{  The SPAed transpose for qutrit states has been performed with an instance of qutrit state \cite{Lim:2012aa}. (a) A randomly prepared state is characterized by quantum state tomography. (b) The ideal transpose, non-physical, is shown. The ideal SPAed transpose shows the transformation in (d), which has been demonstrated in experiment in (c). The experimental results are reported as a proof-of-principle demonstration.} 
\label{fig:qutritTexp}
\end{center}
\end{figure}

In Ref. \cite{Lim:2012aa}, the aforementioned construction is applied to qutrit SIC states and its experimental proof-of-principle demonstration has been presented. The measure-and-prepare scheme in Eq. (\ref{eq:sicT}) for $d=3$ is performed in experiment with single-photon states with spatial and polarization degrees of freedom. The following nine SIC states are generated with a fiducial state and group action and applied to the implementation, \cite{Renes:2004aa} 
\bea
&& |v_1\rangle =\frac{1}{\sqrt{2}}  \left( \begin{array}{ccc} 1 \\ w \\ 0 \end{array} \right),~ 
|v_2\rangle =\frac{1}{\sqrt{2}}  \left( \begin{array}{ccc} 1 \\ w^2 \\ 0 \end{array} \right),~
|v_3\rangle =\frac{1}{\sqrt{2}}  \left( \begin{array}{ccc} 1 \\ w^3 \\ 0 \end{array} \right),~ \nonumber \\
&&
|v_4\rangle =\frac{1}{\sqrt{2}}  \left( \begin{array}{ccc} 0 \\ 1 \\ w \end{array} \right), ~
|v_5\rangle =\frac{1}{\sqrt{2}}  \left( \begin{array}{ccc} 0 \\ 1 \\ w^2 \end{array} \right),~
|v_6\rangle =\frac{1}{\sqrt{2}}  \left( \begin{array}{ccc} 0 \\ 1 \\ w^3 \end{array} \right), \nonumber\\ 
&&
|v_7\rangle =\frac{1}{\sqrt{2}}  \left( \begin{array}{ccc} w \\ 0 \\ 1 \end{array} \right),~
|v_8\rangle =\frac{1}{\sqrt{2}}  \left( \begin{array}{ccc} w^2 \\ 0 \\ 1 \end{array} \right),~
|v_9\rangle =\frac{1}{\sqrt{2}}  \left( \begin{array}{ccc} w^3 \\ 0 \\ 1 \end{array} \right), \nonumber
\eea
where $w = e^{2\pi i /3}$. These states are prepared in a single photon's polarization and path degrees of freedom as follows. For a single photon having two paths, say $a$ and $b$, three computational basis are identified as $|0\rangle = |a,H\rangle$, $|1\rangle = |a,V\rangle$, and $|2\rangle = |b,H\rangle$. The SPAed map has been implemented as, for a qutrit state $\rho$,
\bea
\widetilde{T} [\rho] = \sum_{k=1}^9 \tr[\frac{1}{3} | v_{k}^*\rangle \langle v_{k}^*| \rho ] | v_k \rangle \langle v_k | \nonumber 
\eea
with the nine SIC states, in a similar way in Fig. \ref{fig:qubitTsetup} (A) as a proof-of-principle demonstration. The experiment has reported that, for an instance of qutrit state, the SPAed transpose is performed with state fidelity $\approx 99.9\%$, and the process tomography shows the gate fidelity $\approx 98.2\%$ \cite{Lim:2012aa}.

\subsection{Partial transpose on two-qubit states}

\begin{figure}
\begin{center}  
\includegraphics[width= 10cm]{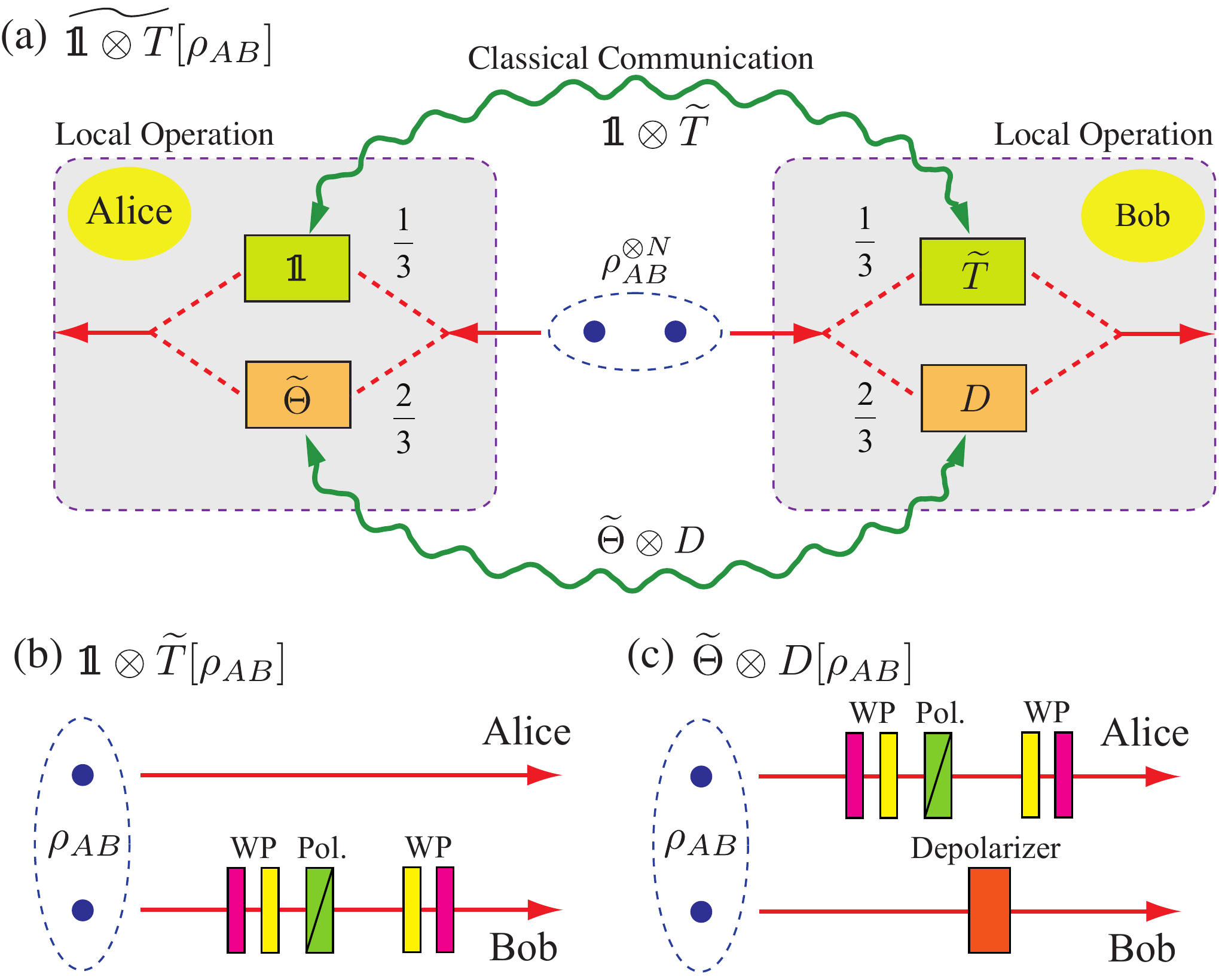}
\caption{ An LOCC protocol for the SPAed partial transpose has been implemented for two photon states with polarization degrees of freedom \cite{PTexpBae}. It is composed of two quantum channels in (b) and (c), both of which are local operations.  } 
\label{fig:PTsetup}
\end{center}
\end{figure}

For further applications, as well as towards direct detection of entanglement with SPAed maps \cite{Horodecki:2002aa} to be discussed in the next section \ref{section:applications}, it is essential to implement the SPAed partial transpose $\widetilde{ \i \otimes T}$, see also Eq. (\ref{eq:spa-ent}). In fact, the SPAed partial transpose is entanglement-breaking for two-qubit states \cite{Fiurasek:2002aa} and also for higher dimensions in general \cite{SPAconjecture}. Therefore, the SPAed partial transpose can be in general implemented in a measure-and-prepare scheme. It is left to devise a measure-and-prepare protocol for the practical implementation.

For the purpose, we refer to the LOCC decomposition given in Eq. (\ref{eq:spaLOCC}) \cite{Alves:2003aa}. The SPAed map on bipartite systems can be performed by SPAed maps on local systems. In particular, applying to the partial transpose for two-qubit states, we have the following resulting state,
\bea
\widetilde{\i \otimes T } [\rho] = \frac{1}{3} ( \i \otimes \widetilde{T}) [\rho] + \frac{2}{3} (\widetilde{\Theta} \otimes D) [\rho] \label{eq:spaIT}
\eea
where $D$ denotes the depolarization map for qubit states and $\widetilde{\Theta}$ is the SPAed inversion map in Eq. (\ref{eq:SPAtheta}). Recall that $\widetilde{T}$ is entanglement-breaking, as well as the depolarization map.

\begin{figure}
\begin{center}  
\includegraphics[width= 10cm]{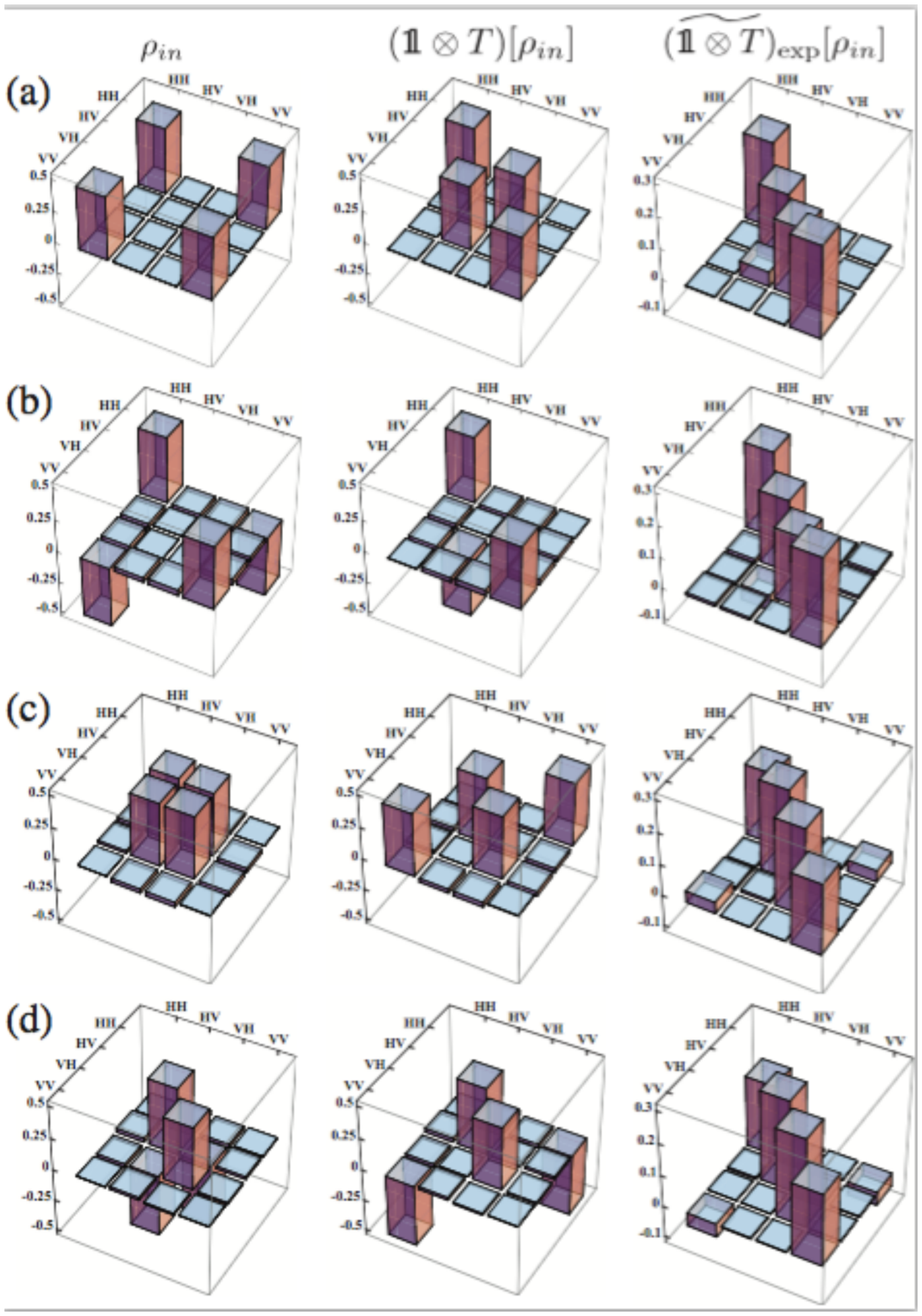}
\caption{ The SPAed partial transpose is applied to four Bell states in experiment \cite{PTexpBae}. (a) and (b) show the approximate partial transpose for $|\phi^{\pm}\rangle = (|00\rangle \pm |11\rangle)/\sqrt{2}$ and $|\psi^{\pm}\rangle = (|01\rangle \pm |10\rangle)/\sqrt{2}$. The second row shows the theoretical computation of the partial transpose for four Bell states, which can be compared to the third row that shows the experimental results of the SPAed partial transpose for the states.  } 
\label{fig:PTqst}
\end{center}
\end{figure}

SPA to the inversion map $\widetilde{\Theta}$ is also entanglement-breaking. One can find that the CJ operator is given by,
\bea
\chi_{\widetilde{\Theta}} = 
 (\ido \otimes Y) \chi_{\widetilde{T}} (\ido \otimes Y) \nonumber
\eea
where $\chi_{\widetilde{T}}$ in Eq. (\ref{eq:tch}) is separable and $Y$ denotes the Pauli matrix. That is, the CJ operator $\chi_{\widetilde{\Theta}}$ is equivalent to the separable state $\chi_{\widetilde{T}}$ up to local unitary $Y$ and hence, separable. This also constructs the prepare-and-measure scheme as follows, 
\bea
\widetilde{\Theta} [\rho] = \sum_{k=1}^4 \tr[\frac{1}{2} | v_{k}^*\rangle \langle v_{k}^*| \rho] |w_k\rangle \langle w_k | \nonumber
\eea
where $\{| v_k\rangle \}_{k=1}^4$ are SIC states, e.g. in Eq. (\ref{eq:tetrastate}) and we have put $|w_k\rangle = Y |v_k\rangle$ for $k=1,2,3,4$. Therefore, the SPAed partial transpose in Eq. (\ref{eq:spaIT}) can be performed by applying two quantum channels $\i \otimes \widetilde{T}$ and $\widetilde{\Theta}\otimes D$ with probabilities $1/3$ and $2/3$, respectively, where both are local operations. The implementation scheme is shown in Fig. \ref{fig:PTsetup} (a). 

In Ref. \cite{PTexpBae}, the SPAed partial transpose has been realized for two photon states with polarization degrees of freedom, see Fig. \ref{fig:PTsetup}. For for the channel $\i \otimes \widetilde{T}$, the scheme for the SPAed transpose in Fig. \ref{fig:qubitTsetup} (A) is exploited. The experimental realization of the SPAed partial transpose has been performed for four Bell states. In Fig. \ref{fig:PTqst}, the experimental results are shown and compared with theoretical calculations. The experiment results show high fidelities about $99\%$ between ideal and experimental SPAed transpose operations.

Finally, we note that the results shown here for the SPAed partial transpose can also be applied to other SPAed maps $\widetilde{\i\otimes \Lambda}$ with further efforts of finding a measure-and-prepare scheme for the SPAed map $\widetilde{ \Lambda}$ on a local system.

\subsection{Quantifying implementation of noisy quantum channels}

In the previous subsections, we have considered implementation of SPAed transpose and SPAed partial transpose with photon polarization qubits. The experimental results in Refs. \cite{Lim:2011aa, PTexpBae, Lim:2012aa} have reported about $99\%$ in both state fidelity and gate fidelity. On the other hand, as a similar consideration, the experimental realization of the UNOT operation in Ref. \cite{De-Martini:2002aa} has achieved about $95\%$ \footnote{The optimal fidelity between the ideal UNOT and approximate, and physical, UNOT is given by $2/3\approx 0.666$ \cite{Buzek:1999aa}. The experimental realization has reported a fidelityabout $0.63$ \cite{De-Martini:2002aa}. Then, one can find that the experimental realization has fidelity $95\%$ with respect to the optimal operation.}. All these show high fidelities in common. 


In Ref. \cite{arxivbae}, it has been discussed to answer the question how the high values in fidelities could be obtained in those experiments. The conclusion is drawn that, simply saying, SPAed operations are so noisy that they are close to a complete depolarization. One can actually find a SPAed map is a quantum channel around the complete depolarization, which is not difficult to realize experimentally. Let us show the detailed explanation in the following. 

Let us begin with quantum state fidelity that is useful when estimating similarity of two quantum states. For two states $\rho, \sigma \in S(\H)$, the fidelity is given by 
\bea
F(\rho,\sigma) = \tr \sqrt{\sqrt{\rho} \sigma \sqrt{\rho}} = \| \sqrt{\rho} \sqrt{\sigma} \|_1, \nonumber
\eea
which is called the Uhlmann fidelity \cite{Uhlmann:1976aa}. Let $D_{\mathrm{tr}}$ denote the trace distance \footnote{The trace distance is given by the $L_1$-norm, $D_{\mathrm{tr}} (\rho,\sigma) = \| \rho - \sigma\|_1 /2 $}. Then, the state fidelity is in fact equal to the reciprocal of the trace distance $1-D_{\mathrm{tr}}(\rho,\sigma)$ when either of $\rho$ or $\sigma$ is a pure state. In general, we have $F (\rho,\sigma) \geq 1-D_{\mathrm{tr}} (\rho,\sigma)$. Based on the state fidelity, gate fidelities can be introduced to compare two quantum operations, $\E_1$ and $\E_2$. One can define two measures, the average gate fidelity $F_{\mathrm{ave}}$ and the worst-case fidelity $F_{\mathrm{w}}$, in the following way,
\bea
F_{\mathrm{ave}} (\E_1 ,\E_2) & = & \int d\rho F( \E_1 (\rho) , \E_{2} (\rho)),~\mathrm{and},\nonumber \\
F_{\mathrm{w}} (\E_1 ,\E_2) & = & \min_{\rho} F(\E_1 (\rho) , \E_{2} (\rho) ).\nonumber
\eea
These gate fidelities have operational meanings. 

The gate fidelities can be applied to comparing ideal and experimental operations. Let $\E$ denote the ideal quantum operation that experimentalists aim to realize in experiment, and let $\E_{\mathrm{exp}}$ denote an experimental realization of the ideal one. Then, an implementation by $\E_{\mathrm{exp}}$ can be quantified by the gate fidelities $F_{\mathrm{ave}} (\E, \E_{\mathrm{exp}})$ or $F_{\mathrm{w}} (\E, \E_{\mathrm{exp}})$. Once it happens $F_{\mathrm{ave}}$ =1 or $F_{\mathrm{w}}=1$, one can understand the agreement of experimental implementation and the desired operation with certainty.

\begin{figure}
\begin{center}  
\includegraphics[width= 12cm]{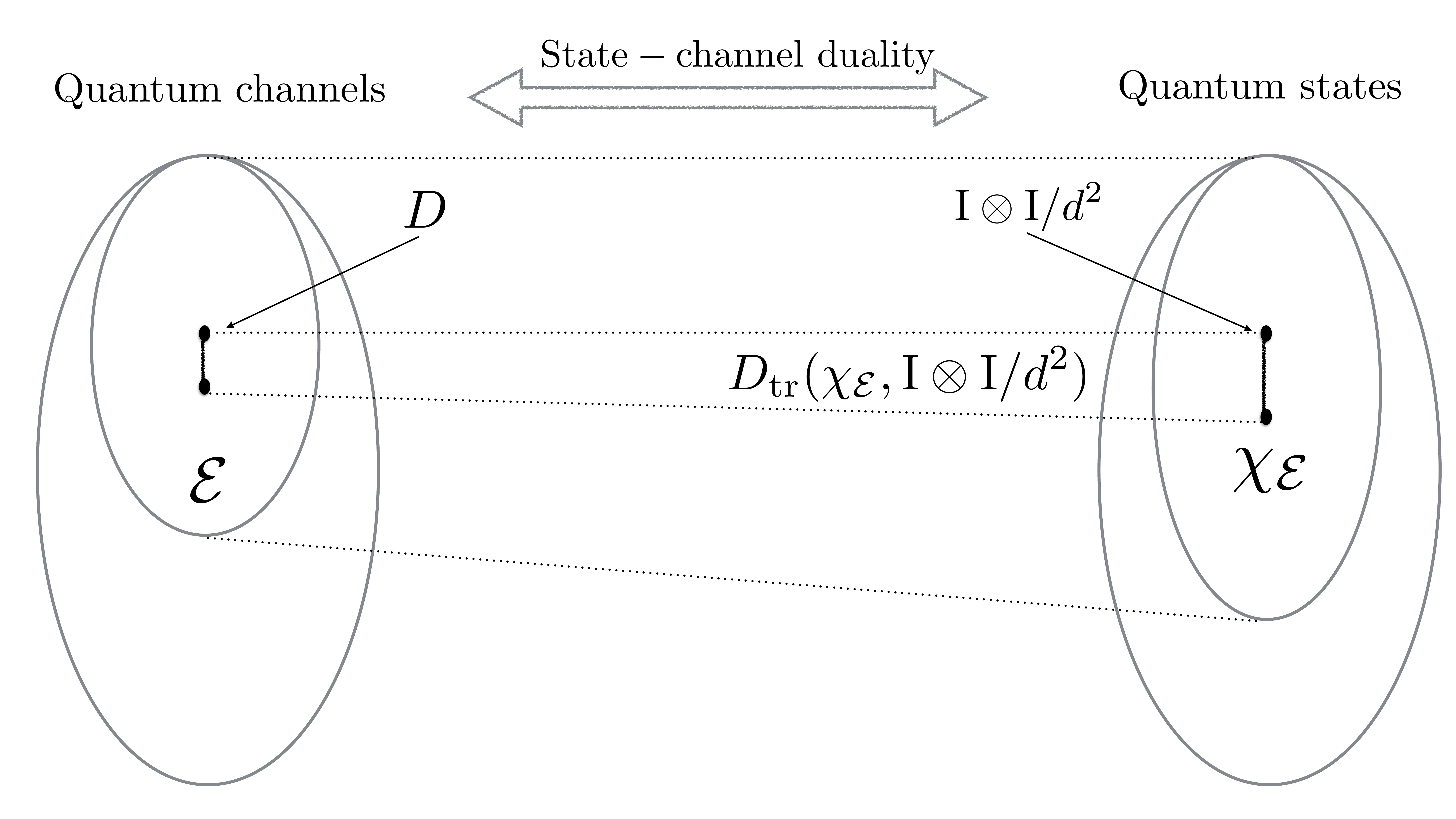}
\caption{The trace distance of CJ operators of quantum channels are depicted, which also provides lower bounds to gate fidelities, see Eq. (\ref{eq:fdistance}). An entanglement-breaking channel $\E$ is close to the complete contraction map $D$ in the $L_1$ distance of their CJ operators. Even the complete depolarization is realized, it has a high gate fidelity with respect to the desired one $\E$ since they are close enough to have a high lower bound to a gate fidelity.} 
\label{fig:fdistance}
\end{center}
\end{figure}

The key observation is that the operation desired to implement when applying SPA would not be in the case when the unit-valued fidelity is possible, since SPA is applied to impossible operations. For instance, no quantum operation can achieve the transpose map in experiment. Recall that the optimal approximate UNOT operation, unitarily equivalent to the transpose, can achieve gate fidelities $2/3$ \cite{Buzek:1999aa}. The fidelity estimates how similar the quantum operation and the impossible one is. Once the approximate map is realized, it makes sense to compare the implementation with the approximate one for the quantification of the performed experiment, in which gate fidelities are upper bounded by the unit.

All these can be seen in terms of CJ operators. The SPA conjecture addresses that CJ operators to optimal positive maps are separable states. When positive maps do not satisfy the conjecture, their CJ operators are sufficiently close to separable states. In Ref. \cite{arxivbae}, it is shown that for an entanglement-breaking channel $\E$ the gate fidelities are lower bounded as,
\bea
F(\E_{\mathrm{exp}},\E ) \geq 1- d D_{\mathrm{tr}}( \chi_{ \E_{\mathrm{exp}} },\chi_{\E}) \label{eq:fdistance}
\eea
while it is upper bounded by the unit. Note that $\chi_{\E}$ is separable or close to the separable states in the $L_1$ norm, so is $\chi_{\E_{\mathrm{exp}}}$. Thus, the value $D_{\mathrm{tr}}( \chi_{ \E_{\mathrm{exp}} },\chi_{\E})$ is small enough as it is at most a distance within the set of separable states. This means that, even in worst cases that an arbitrary entanglement-breaking channel $\E_{\mathrm{exp}}$ is realized as an implementation of desired one $\E$ , fidelities are at least as high as the lower bound. For instance, for the case of the SPAed transpose map for qubit states, the lower bound is immediately about $98\%$ \cite{arxivbae}.  We remark that high lower bounds are, not because of an experimentally realized operation, but due to the entanglement property of the CJ operator of a quantum operation that is desired to realize.


\setcounter{footnote}{0}
\section{ Applications to Entanglement Detection}
\markboth{\sc Applications to Entanglement Detection}{}
\label{section:applications}

Since SPA was proposed \cite{Horodecki:2003ab}, the first application has been found in entanglement detection \cite{Horodecki:2002aa}. SPA-based entanglement detection involves in more complicate steps than the other with EWs. There has also been simplification of the original scheme with the SPA conjecture in Ref. \cite{SPAconjecture}. In this section, we overview the recent progress in entanglement detection with SPA. We also address the usefulness of SPAed EWs for detecting entangled states under weaker assumptions.

\subsection{Entanglement detection in practice }

Let us begin by describing the scenario of entanglement detection. We discuss different strategies of detecting entangled states and compare advantages and disadvantages. Depending on whether given quantum states have been identified, or not, one can make different approaches to entanglement detection. The scenario often appeared in quantum information tasks is that a bipartite quantum state, denoted by $\rho_{AB} \in S (\H\otimes \H)$, is repeatedly generated from a device that after $N$ repetitions the collected states are given by $\rho_{AB}^{\otimes N}$. 


{\bf Theoretical tools after state tomography.} Quantum state tomography followed by a theoretical analysis can generally decide if state $\rho$ is entangled or separable. When the dimensions of underlying Hilbert space are known but the quantum state, a complete measurement can be obtained and applied to identifying quantum states. Note that SIC POVMs can construct a minimal number of detectors for state tomography, as long as their existence is known. Since the number of SIC POVMs is given by $d^2$ in a $d$-dimensional Hilbert space, tomography for $n$-partite quantum states requires about $d^{2n}$ detectors at least in measurement, denoted by $O(d^{2n})$. That is, the number of detectors actually grows exponentially, which may make it infeasible to perform tomography for sufficient large systems. 

Once state $\rho_{AB}$ is identified, one can apply theoretical tools to determine if the state is entangled, or separable. For instance, as mentioned in the subsection \ref{subsection:entdetection}, the complete set of positive maps can detect all entangled states. Recall that a quantum state $\rho_{AB} \in S (\H\otimes \H)$ is entangled if and only if there is a positive map $\Lambda$ such that $(\i \otimes \Lambda) [\rho] \ngeq 0$. In this case, the difficulty lies at the fact that the structure of positive maps is generally not known, which has been idefined as a mathematically challenging problem. There are known positive maps that can be applied, such as the partial transpose, Choi maps, Breuer-Hall maps \cite{Breuer:2006aa, Hall:2006aa}, etc., and may suffice to detect some classes of entangled states. 

\begin{figure}
\begin{center}  
\includegraphics[width= 10cm]{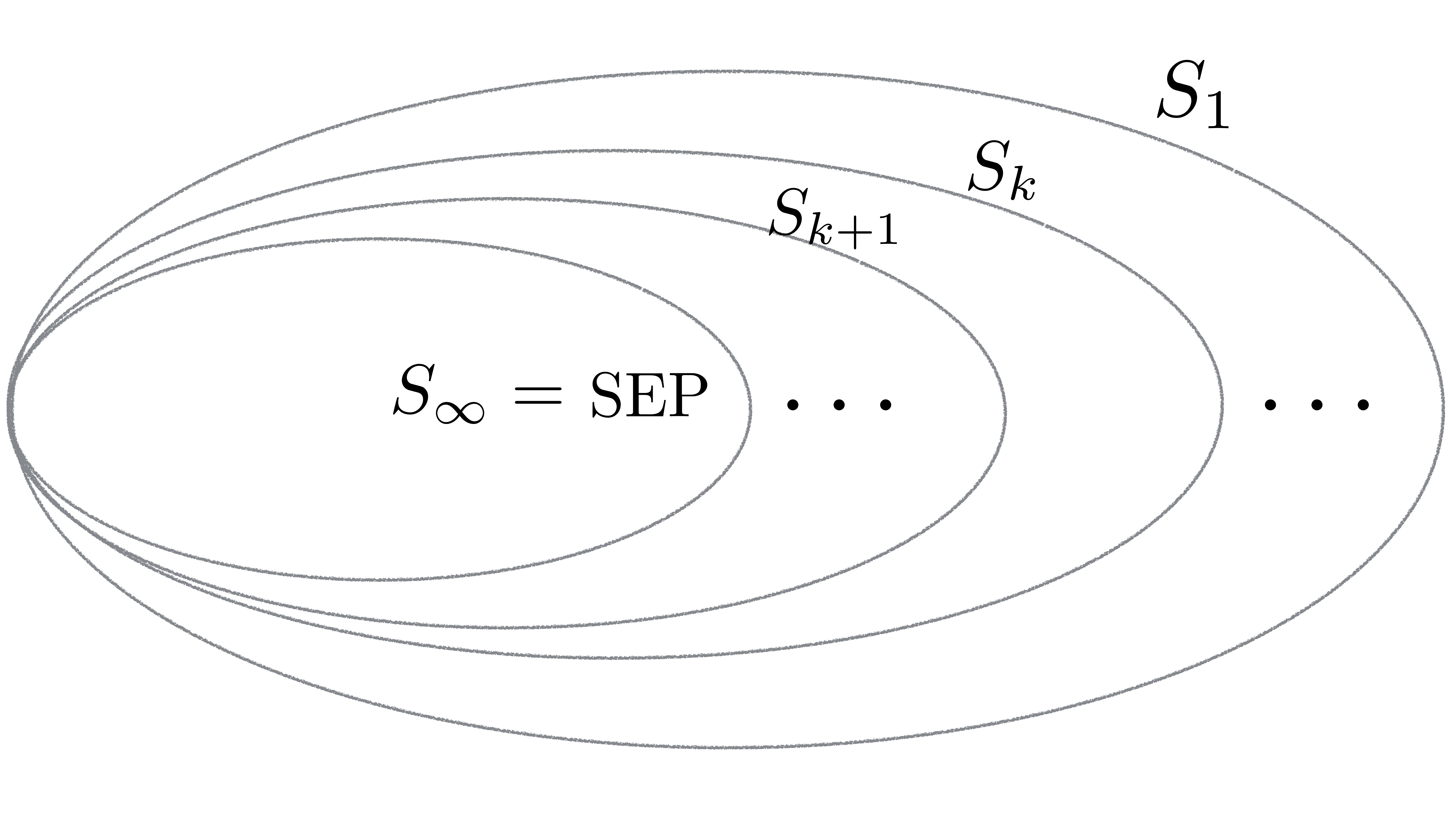}
\caption{ The set of $k$-extendiable states is denoted by $S_k$, and the natural relation of inclusion is shown. The hierarchy can be exploited to making numerical approaches to detecting entangled states. }
\label{fig:hierarchy}
\end{center}
\end{figure}

On the other hand, numerical methods can be applied to the separability problem by exploiting the semidefinite program (SDP). The approach is based on the so-called extendibility of quantum states, or equivalently monogamous property of entangled states. That is, entanglement cannot be shared by arbitrarily many parties and infinitely shareable states are only separable states. A bipartite state $\rho_{AB} \in S(\H\otimes \H)$ is called $N$-extendible if there exists $(N+1)$-partite state $\rho_{AB_1 B_2 \cdots B_N}$ such that 
\begin{enumerate}
\item the $(N+1)$-partite state is permutationally invariant over parties $B_1 \cdots B_k$, i.e. $\mathrm{P}_{B_{1}B_2\cdots B_k} \rho_{AB_1 B_2 \cdots B_k}  \mathrm{P}_{B_{1}B_2\cdots B_k}  = \rho_{AB_1 B_2 \cdots B_k}$ under permutations $ \mathrm{P}_{B_{1}B_2\cdots B_k} $ over $B_1 B_2 \cdots B_k$, and \\
\item $\rho_{AB_i} = \rho_{AB_j}$ for all $0\leq i,j \leq N$ where $\rho_{AB_i}$ is the reduced state from $\rho_{AB_1 B_2 \cdots B_N}$ after tracing out all but $AB_i$. \end{enumerate}
Denoted the set of $k$-extensible bipartite states by $S_k$, there exists a natural hierarchy, see Fig. (\ref{fig:hierarchy}),
\bea
S_{\infty} \subset \cdots \subset  S_{k+1} \subset S_{k} \subset \cdots \subset S_2 \subset S_1. \nonumber
\eea
Then, it turns out that the only bipartite state $\rho_{AB}$ that admits $\infty$-extendibility is a separable state \cite{Werner:1989ab}. This means that a state is entangled if and only if there exists a finite number $k$ such that the state admits only $k$-extension. 

The extendibility is constrained with a feasible constraint, positive-partial-transpose (PPT) extension: state $\rho_{AB}$ is $N$-PPT-extendible if its extension $\rho_{AB_1 \cdots B_N}$ is a PPT state in all bipartitions $ N/2+1 :N/2$. In fact, infinitely PPT extendible states coincide to the set of separable states. The hierarchy has been exploited to compose a numerical program of detecting entangled states with SDP \cite{PhysRevLett.90.157903, PhysRevLett.103.160404, PhysRevLett.109.160502}. 

We conclude quantum state tomography followed by known theoretical tools as an approach to detecting entangled states. The drawbacks are, first of all, the experimental cost for quantum state tomography where the required resources increase exponentially. Then, theoretical tools are also limited in that i) the structure of positive maps remains open and ii) the numerical approach is useful but generally considered to be intractable as it is in NP-Hard.

{\bf Entanglement witnesses.} With EWs, one can detect entangled states even before quantum states are verified by tomography, which is termed as {\it direct detection} of entanglement. We recall that EWs are observables such that they have non-negative expectation values for all separable states, see Eq. (\ref{eq:EW}), and negative ones for some entangled states. Since EWs are observables, they can be directly realized in experiment. 

Since an operator corresponding to an EW is Hermitian, it can be factorized into projections, or POVMs in general, such that $W = \sum_{i} c_i P_i$ where $P_i \geq 0$ denoting a POVM element corresponds to a detector, i.e., a detector is described by a positive operator. Then, expectation value $\tr[W\rho]$ can be obtained in experiment by finding probabilities from detection events. Recall that, for state $\rho$, the probability that a detector described by $P_i$ shows an event "click" is given by $\tr[P_i\rho]$. Then, expectation $\tr[W\rho]$ for some state $\rho$ can be obtained by finding the values $\tr[P_i \rho]$ experimentally and combining them with coefficients $c_i$, i.e. $\sum_i c_i \tr[ P_i \rho]$. 

In general, EWs can also be factorized into local observables \cite{Guhne:2003aa}, so that they can be applied to a scenario where entanglement is detected by two parties far in distance, i.e.,
\bea
W = \sum_{i} c_{i} O_{i}^{(A)} \otimes O_{i}^{(B)} \label{eq:localEW}
\eea
with local observables of $O_{i}^{(A)}$ and $O_{i}^{(B)}$. Although factorization into local observables is not necessary for detecting entangled states, we here consider local observables, i.e. without joint measurement, to discuss the connection to the experimental cost of quantum state tomography that applies local measurement only. In fact, joint measurement requires additional experimental costs of making quantum systems interact one another. 

Then, an observable can also be decomposed into POVMs, denoted as $O_i = \sum_{j} x_{ij} P_{j}$ with $\{x_i \}$ are some constants and $\{P_i \}$ are POVM elements. In this way, an EW in Eq. (\ref{eq:localEW}) can be described with local POVMs as
\bea
W = \sum_{ij} \widetilde{c}_{ij} P_{i}^{(A)} \otimes P_{j}^{(B)} \label{eq:LEW}
\eea
with some constants $\{\widetilde{c}_{ij}\}$. Since each POVM corresponds to a description of a detector, the expectation value $\tr[W\rho]$ for state $\rho$ is found by estimating the quantity in the following
\bea
\sum_{ij} \widetilde{c}_{ij} p(ij | \rho)~~\mathrm{where} ~~ p( ij | \rho) = \tr[P_{i}^{(A)} \otimes P_{j}^{(B)} \rho].  \label{eq:detp}
\eea
As this corresponds to the expectation $\tr[W\rho]$, a state $\rho$ must be entangled if it appears that $\sum_{ij} \widetilde{c}_{ij} p(ij | \rho)<0$. We here conclude that EWs as a method of direct detection of entangled states, and also refer to an excellent review on EWs for further considerations \cite{Guhne:2009aa}. Now, crucial is the number of detectors in Eq. (\ref{eq:LEW}) compared to quantum state tomography, see the discussion in the below.

{\bf Comparison.} Let us summarize advantages and disadvantages of the aforementioned approaches, quantum state tomography followed by theoretical methods, and EWs. Despite the computational complexity of the separability problem, one can find that once a quantum state is identified, known theoretical methods mentioned above such as positive maps or SDP work sufficiently well for practical purposes, in particular for low dimensional systems. For instance, the partial transpose criteria can characterize all two-qubit separable states: the transpose map can determine whether a two-qubit state is entangled or not. One can, however, notice that quantum state tomography is in fact an expensive process. To perform tomography for an $n$-partite state in Hilbert space $\H_{d_1} \otimes \H_{d_2}  \otimes \cdots \otimes \H_{d_n}$, where each subscript denotes the dimension of the space, the number of detectors required for tomographically complete measurement is at least given by $d_{1}^2  d_{2}^2 \cdots d_{N}^2 $. Once measurement outcomes are collected, it also takes significant amount of time for a numerical algorithm to reconstruct a quantum state. 

EWs can bypass the step of quantum state tomography for detecting entangled states. Given an EW, as soon as it is observed $\sum_{ij} \widetilde{c}_{ij} p(ij | \rho) <0$ in Eq. (\ref{eq:detp}), that confirms $\tr[W \rho] <0$, the state must be entangled. This is particularly useful when entanglement detection is more significant than verification of quantum states. The disadvantage, however, exists in a low efficiency of detecting entangled states. Simply saying, even in the simplest case of two-qubit states, there is no single EW that can detect all entangled states \footnote{This can be easily seen as follows. Let $W$ denote an EW that detects both $\rho_1 = |\phi_1\rangle \langle \phi_1 |$ and $\rho_2 = |\phi_2 \rangle \langle \phi_2 |$. That is, we have $\tr[W  \rho_1] < 0$ and $\tr[W  \rho_2]$. A contradiction is then drawn for the state 
$\sigma = (\rho_1 + \rho_2)/2$, which is a separable state. However, it holds that $\tr[ W \sigma] = (\tr[ W\rho_1  ] + \tr[W \rho_2] )/2 <0$, that contradicts to the assumption that $W$ is an EW.  }. We recall the relation that a witness $W$ is derived from a positive map $\Lambda$, i.e., $W =(\i \otimes \Lambda) [Q]$ by choosing some positive operator $Q$ from Eq. (\ref{eq:dual}). It is clear that a single positive map can detect more of entangled states than an EW.

The question arising when applying EWs in practice is in fact the experimental cost for realizing and processing EWs. Experimentalists can simply ask if detectors used for EWs can also perform state tomography. This means that EWs and quantum state tomography can be performed with the same experimental costs and are different only in the classical processing, i.e., processing of measurement outcomes to conclude if given quantum states are entangled. If tomography is performed to verify a quantum state, theoretical tools can be applied to dectermine if it is entangled or separable. If an EW is constructed and expectation in Eq. (\ref{eq:detp}) is estimated, it detects a fraction of entangled states. 

EWs take their own advantage when their measurement  is not sufficient to perform tomography, i.e., when they form a tomographically incomplete measurement. This then asks the non-trivial problem of minimizing experimental resources required for EWs. That is, the number of POVMs in Eq. (\ref{eq:LEW}) is asked to be minimized. In the following subsections, we revisit two schemes of detecting entangled states, i) direct application of positive maps in experiment, and ii) minimal resources to realize EWs. 

\subsection{Entanglement detection with structural physical approximation}

In Ref. \cite{Horodecki:2002aa}, a method of direct detection of entanglement has been proposed with explicit application of positive maps. For a map $\i\otimes \Lambda$ that detects entangled states, SPA leads to the following, 
\bea
\widetilde{\i\otimes \Lambda} = (1-p^*)\i\otimes \Lambda + p^* D\otimes D \nonumber 
\eea
with the minimal $p^*$ such that the resulting map is CP, see Eq. (\ref{eq:spa-ent}) for the parameters. If for unknown state $\rho$ the SPAed map $\widetilde{\i \otimes \Lambda}[\rho]$ may show eigenvalues less than $p^*/d^2$, then one can conlclude that the state is entangled. This is because, for state $\rho$, the minimum eigenvalue of $(\i\otimes\Lambda)[\rho]$ denoted by $ \lambda $ is related to that of $(\widetilde{\i \otimes \Lambda]} )[\rho]$ by $\widetilde{\lambda}$, as
\bea
\widetilde{\lambda} = (1-p^*) \lambda + \frac{p^{*}}{d^2}. \label{eq:eig-relation}
\eea 
To see this, let us suppose that $Q>0$ is a rank-one projector such that $\lambda = \min_Q \tr[Q~(\i\otimes \Lambda)[\rho] ]$. Then, suppose that $Q^{'}$ is a rank-one projector having $\widetilde{\lambda} = \min_{Q^{'}}  \tr[Q^{'} ~ \widetilde{\i\otimes \Lambda} [\rho] ]$. Then, the eigenvalue $\widetilde{\lambda}$ is given by $(1-p^*) \min_{Q^{'}} \tr[Q^{'}~(\i\otimes \Lambda)[\rho] ] + p^* / d^2$, that shows the relation in Eq. (\ref{eq:eig-relation}). 

From the relation in Eq. (\ref{eq:eig-relation}), if $\rho$ is detected by $\i\otimes \Lambda$ it is also detected by $\widetilde{\i\otimes \Lambda}$, since the condition $\lambda < 0$ means $\widetilde{\lambda} < p^* / d^2$. Conversely, for some state $\rho^{'}$ detected by $\widetilde{\i\otimes \Lambda}$, it is also detected by $\i\otimes \Lambda$ since the condition $\widetilde{\lambda} < p^{*} / d^{2}$ implies that $\lambda < 0$. Therefore, we conclude that a state $\rho$ is entangled if 
\bea
\min_{Q}\tr[Q ~\widetilde{\i\otimes \Lambda} [\rho] ] < \frac{p^{*}}{ d^2}. \label{eq:detection}
\eea
Moreover, there is no loss in the capabilities of detecting entangled states under the condition from $\lambda <0$ to $\widetilde{\lambda}<p^{*}/d^2$ by SPA.

\begin{figure}
\begin{center}  
\includegraphics[width= 13cm]{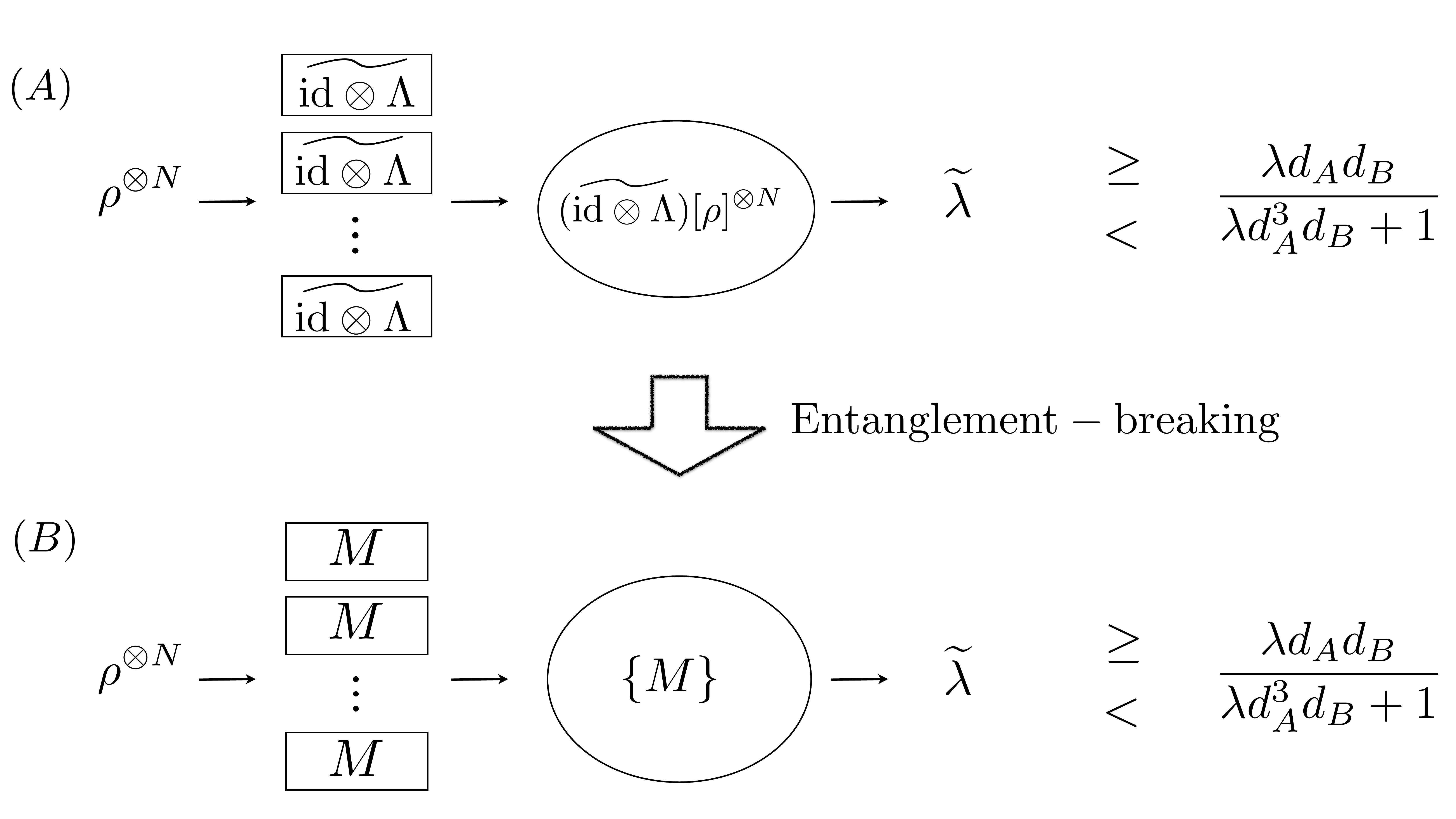}
\caption{ (A) When SPA is applied to entanglement detection, the SPAed map $\widetilde{\i \otimes \Lambda}$ for a positive map $\Lambda$ is applied to copies of given states and the resulting states states are collected \cite{Horodecki:2002aa}. The spectrum estimation, e.g. in \cite{Keyl:2001aa}, is applied to find the minimum eigenvalue $\widetilde{\lambda} = \min_{Q>0} \tr[Q  [ \widetilde{\i\otimes\Lambda} ](\rho) ]$. If the eigenvalue is smaller than $\lambda d_A d_B / (\lambda d_{A}^{3} d_B +1 )$ where $\lambda = \min_{Q>0} \tr[Q  [  \i\otimes\Lambda  ](\rho) ]$. (B) When the SPAed map $\widetilde{\i \otimes \Lambda}$ is entanglement-breaking, the operation can be performed by a measure-and-prepare scheme. Then, the spectrum estimation for the collected states can be replaced by a post-processing on measurement outcomes.}
\label{fig:spectrum}
\end{center}
\end{figure}

Having derived the detection condition in Eq. (\ref{eq:detection}), we are now left with estimation of minimum eigenvalues. In Ref. \cite{Horodecki:2002aa}, the detection scheme refers to the spectrum estimation in Ref. \cite{Keyl:2001aa}, see Fig. \ref{fig:spectrum}, where measurement on the symmetric subspace is applied, that is in fact joint measurement. This asks quantum memory to store quantum systems for a while, that is however experimentally challenging. 

In Ref. \cite{SPAconjecture}, the SPAed map $\widetilde{ \i \otimes T}$ is in general entanglement-breaking, meaning that the map $\widetilde{ \i\otimes T}$ can be implemented by measurement and preparation of quantum states. Namely, there exists rank-one operators $\{ M_k \}$ and $\{\sigma_k \}$ such that for state $\rho$,
\bea
\widetilde{\i\otimes T} [\rho] = \sum_k \tr[M_k \rho ] \sigma_k. \nonumber
\eea
Since the SPAed map can be realized by a measure-and-prepare scheme, the spectrum estimation scheme can be done from measurement outcomes from correspond POVMs $\{ M_k\}$. This then leads to huge simplification that quantum memory, required for applying joint measurement, is not necessary.

Furthermore, as it is explained in subsection \ref{section:spaexp}, an SPAed map $\widetilde{\i\otimes T}$ can be decomposed into local operations such that it can be realized by an LOCC scheme. Recall that, see Eq. (\ref{eq:spaLOCC}) for detailed parameters and maps,
\bea
\widetilde{\i\otimes \Lambda} = (1-q) \i \otimes \widetilde{\Lambda} + q \widetilde{\Theta} \otimes D_{A\rightarrow B}. \nonumber 
\eea
Note also that in the above, both $\widetilde{\Theta}$ and $D_{A\rightarrow B}$ are entanglement-breaking. For cases where $\widetilde{\Lambda}$ is entanglement-breaking, the SPAed map $\widetilde{\i\otimes \Lambda}$ for a map $\Lambda$ can be realized by local measurement and classical communication, without resort to the requirement of quantum memory. Therefore, it is shown that the detection scheme proposed in Ref. \cite{Horodecki:2002aa} can be implemented by a measure-and-prepare scheme, that is, which is feasible with current technologies. 
 
\subsection{Minimal resources for entanglement detection} 

As it is discussed when EWs take their advantage, the crucial question when applying EWs in practice is the comparison to the complete approach, quantum state tomography followed by theoretical tools. From the conclusion drawn in the above, the goal is now to determine the minimal number of detectors when realizing EWs such that the number is radically less than those of quantum state tomography. We also recall that the number of detectors for tomography increases exponentially with respect to the number of parties. 

In Ref. \cite{EDHOM16}, it has been shown that only $2$ detectors suffice to implement EWs. That is, any EW can be realized with only two detectors, more precisely, two detectors in a Hong-Ou-Mandel (HOM) interferometry. The proposal applies SPAed EWs to the detection scheme. For a witness $W \in \B (\H_d\otimes \H_d)$, its SPAed EW is given by,
\bea
W~~ \longmapsto ~~ \widetilde{W}  = (1-p^* ) W + p^* \frac{\ido \otimes \ido}{d^2}  \label{eq:relationWW}
\eea
Recall that the detection condition for $W$ is given such that state $\rho$ must be entangled if $\tr[W\rho]<0$. From Eq. (\ref{eq:relationWW}), it holds that for state $\rho$
\bea
\tr[W\rho] = \frac{1}{1-p^*} ( \tr[ \widetilde{W} \rho] -\frac{p^*}{d^2})   \label{eq:mod}
\eea
which shows detecting entangled states if $\tr [ \widetilde{W} \rho ]  < p^*/d^2$. One can notice that $\widetilde{W}$ is not only an observable but also a quantum state, i.e., it satisfies $\tr[\widetilde{W}]=1$ and $\widetilde{W}\geq 0$. This motivates one to expect that the quantity $\tr[ \widetilde{W} \rho]$ may be estimated with an interferometry. If it is the case, one does not have to go through the standard steps of implementing EWs, e.g. finding a decomposition of EWs and preparing detectors accordingly, but estimation of a specific parameter that corresponds to the quantity $\tr[ \widetilde{W} \rho]$. When states $\widetilde{W}$ and $\rho$ are given in single photons, the quantity $\tr[ \widetilde{W} \rho]$ is directly related to the probability of coincidence detection in a Hong-Ou-Mandel (HOM) interferometry.

\begin{figure}
\begin{center}  
\includegraphics[width= 10cm]{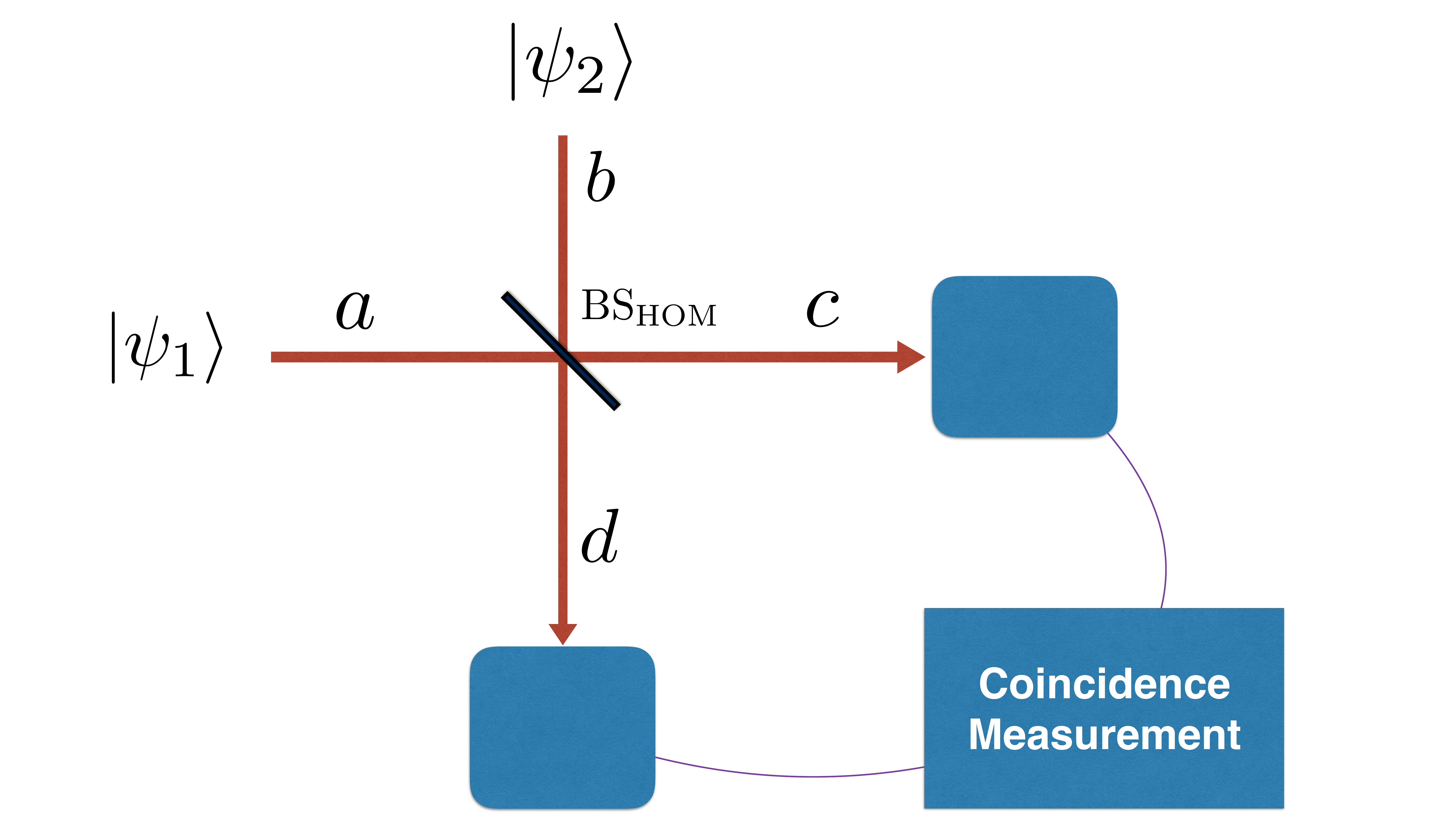}
\caption{  The HOM interferometry is composed of a beam splitter and two detectors (details see text). } 
\label{fig:HOM}
\end{center}
\end{figure}

For convenience, let us suppose that single photons are sent in two input arms of a $50:50$ beamsplitter, see Fig. \ref{fig:HOM} where single photons are sent in the arms $a$ and $b$. If they are indistinguishable, i.e., identical single photons, it only happens that two photons pass the beamsplitter and found in the same mode together, either $c$ or $d$. That is, we have the coincidence probability $p_c = 0$. If two photons are distinguishable, that is, they are prepared in mutually orthogonal states in one of degrees of freedom such as polarization or angular momentum, with probability $1/2$ it happens that one photon is found in $c$ mode and the other in $d$ mode, hence $p_c =1/2$. Therefore, a HOM interferometry shows the interference pattern that has an operational meaning. Denoted by $\sigma_1 = |\psi_1\rangle \langle \psi_1|$ and $\sigma_2 = |\psi_2\rangle\langle \psi_2|$ single-photon states prepared in the input arms, the relation between two states and the coincidence probability is given by 
\bea
\tr[\sigma_1 \sigma_2] = 1-2p_c, ~\mathrm{where}~\sigma_i = |\psi_i \rangle \langle \psi_i |~\mathrm{for}~i=1,2. \label{eq:visibility}
\eea
The relation is valid when mixed states are prepared in input arms. 

Note that the relation in Eq. (\ref{eq:visibility}) works for single-photon states, and two states $\widetilde{W}$ and $\rho$ are bipartite quantum states. We finally incorporate an experimental technique that has been developed recently, the so-called {\it quantum joining}, that enables one to prepare a multipartite quantum state in a single photon's degrees of freedom \cite{Vitelli:2013aa}. It converts degrees of freedom while preserving the overall states, as follows. Suppose that a quantum system has two degrees of freedom $X$ and $Y$, both of which contains two levels for convenience, denoted by $|x_i\rangle$ and $|y_i\rangle$ for $i=0,1$. Then, a bipartite state of systems $A$ and $B$ with a degree of freedom $X$ can be written as follows
\bea
|\psi \rangle_{X_A X_ B} = \sum_{i,j } c_{i,j}  |x_i \rangle_{X_A} |y_j \rangle_{X_B}. \nonumber
\eea
By quantum joining, one can transform the state in the above as follows, 
\bea
|\psi \rangle_{X_A X_ B} \rangle~ ~\longmapsto ~~ | \psi\rangle_{X_A Y_A} = \sum_{i,j } c_{i,j}  |x_i \rangle_{X_A} |y_j \rangle_{Y_A}, \nonumber
\eea
where the resulting state can be described as a four-dimensional state $|\psi\rangle_A$ of a single system $A$. Quantum joining can be applied to quantum systems containing multi-degrees of freedom and has been experimentally demonstrated with single photons' polarization and spatial degrees of freedom \cite{Vitelli:2013aa}, see also the theoretical analysis and structure of quantum joining \cite{Passaro:2013aa}.

\begin{figure}
\begin{center}  
\includegraphics[width= 10cm]{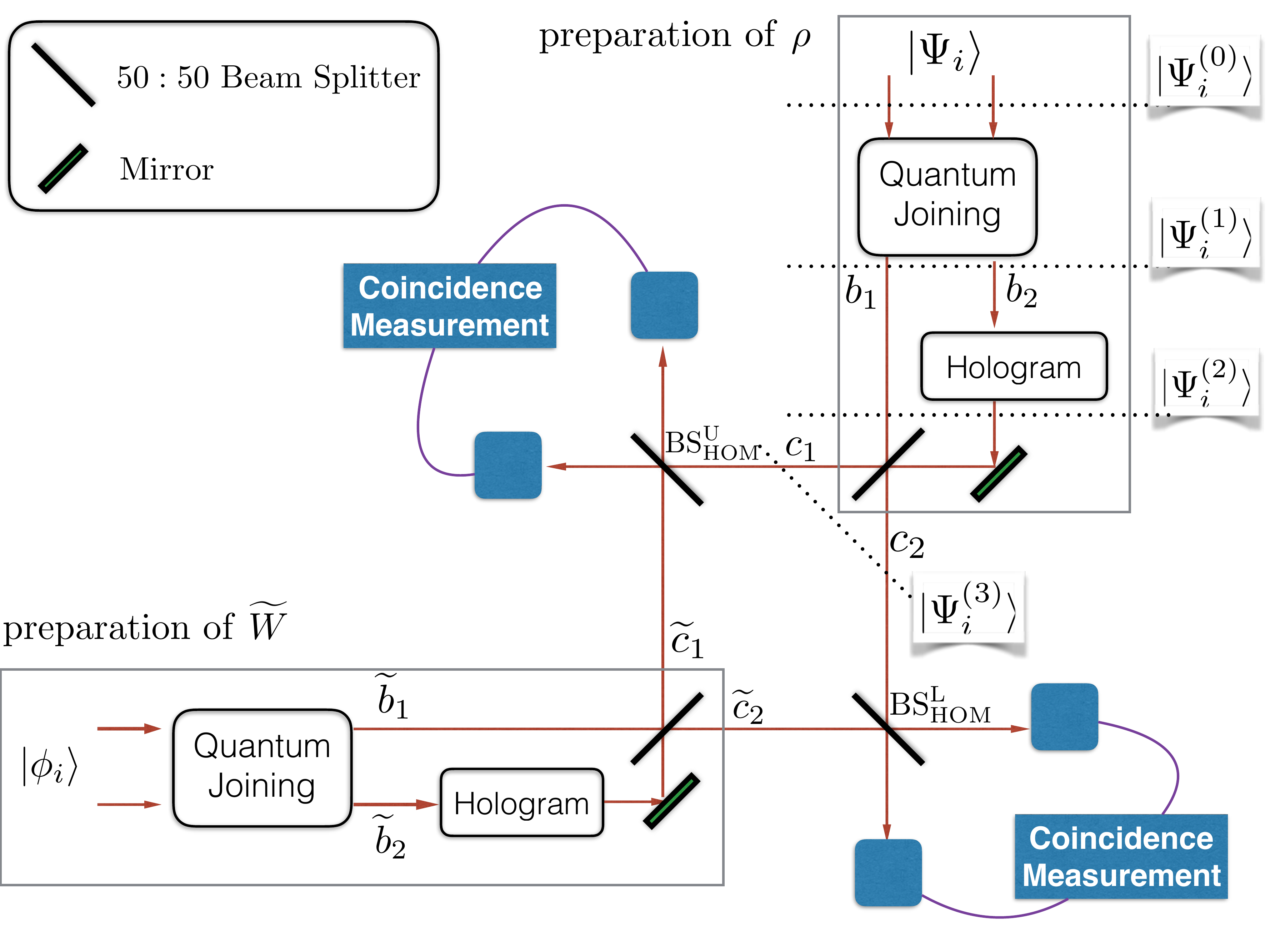}
\caption{ The scheme of entanglement detection with two detectors is shown with applications of quantum joining and OAM degrees of freedome (details see text). In the scheme, the hologram is placed together with a beamsplitter later on to fix the propagation path, either $c_1$ or $c_2$ for state $|\Psi_i\rangle$ and either $\widetilde{c}_1$ or $\widetilde{c}_2$ for state $|\phi_i\rangle$. Then, since the path is given to one of the followings $(c_1, \widetilde{c}_1)$, $(c_1, \widetilde{c}_2)$, $(c_2, \widetilde{c}_1)$, and $(c_2, \widetilde{c}_2)$, two HOM interferometers are applied to enhance the efficiency. } 
\label{fig:2detectorscheme}
\end{center}
\end{figure}

Combining quantum joining and a HOM interferometry, the scheme of detecting entangled states with only two detectors can be devised \cite{EDHOM16}. In Fig. \ref{fig:2detectorscheme}, the coincidence measurement of two states $|\phi_i\rangle $ and $|\Psi_i \rangle$, each of which are bipartite states, is described. Note also that the scheme works for mixed states. Initially, a state $|\Psi_i \rangle$ is prepared in two photons' polarization degree of freedom, i.e., $\{ |H\rangle, |V\rangle\}_A \otimes \{ |H\rangle, |V\rangle\}_B$. Then, the quantum joining scheme in Ref. \cite{Vitelli:2013aa} is applied such that the two-photon state is mapped to a single photon's polarization and spatial degrees of freedom, denoted by $|\Psi_{i}^{(1)}\rangle$. To apply a HOM interferometry later on, the propagation path has to be fixed, i.e., the spatial mode is not yet well fitted to apply a HOM interferometry. To this end, the spatial mode can be converted to orbital-angular-momentum (OAM) degrees of freedom by placing a hologram after quantum joining. Finally, a bipartite state $|\Psi_i\rangle$ is prepared in a single photon's polarization and OAM degrees of freedom, which thus corresponds to a four-dimensional state. The same applies to states $\{|\phi_i\rangle\}$, a decomposition of $\widetilde{W}$, and ends up in a single-photon state. As it is shown in Eq. (\ref{eq:visibility}), one can estimate the coincidence probability $p_c$, by which the quantity $\tr[\widetilde{W}\rho]$ is obtained and applied to determining if given systems are in entangled states. 

The presented scheme can be applied to other physical systems where quantum joining and a HOM interferometry can be applied, see for instance, recent experimental works with atomic states in Ref. \cite{Lopes:2015aa}. Quantum joining corresponds to a unitary transformation \cite{Passaro:2013aa} and thus can be in principle applied to quantum systems that contain multi-degrees of freedom. 

\subsection{Entanglement detection in a measurement-device-independent scenario} 

One of the important direction in quantum information theory is to improve security of quantum cryptography, that is, quantum key distribution. For instance, when quantum channels are noisy, there have been significant efforts to characterize the highest quantum-bit-error-rate that quantum protocols can tolerate \cite{Shor:2000aa, Kraus:2005aa, Chau:2002aa, Gottesman:2003aa, Bae:2007aa, Myhr:2009aa}. Quantum protocols can be improved such that they remain secure when generation of single photons is not certified \cite{Hwang:2003aa, Scarani:2004aa}. In Ref. \cite{Lo:2012aa}, quantum protocols immune to untrusted measurement devices has been proposed, called measurement-device-independent (MDI) protocols.  

In Ref. \cite{Branciard:2013aa}, it has been shown that entanglement detection in an MDI manner turns out to be equivalent to that in the so-called local-operaitons-shared-randomness scenario in Ref. \cite{Buscemi:2012aa}. Remarkably, a systematic method of finding MDI EWs from EWs have been presented, as follows. Let $W$ denote an EW, for which one has to find a decomposition of a witness $W$ in terms of positive operators,  
\bea
W = \sum_{s,t} \beta_{s,t} \tau_s \otimes w_t,~~\mathrm{where}~\tau_s\geq 0~\mathrm{and}~w_t \geq 0. \nonumber
\eea
This precisely corresponds to the decomposition of a given EW in terms of POVMs, see Eq. (\ref{eq:LEW}). Then, an MDI EW can be constructed as
\bea
W_{\mathrm{MDI}} = \sum_{s,t} \beta_{s,t} \tau_{s}^{\top} \otimes w_{t}^{\top},~~\mathrm{where}~\tau_s\geq 0~\mathrm{and}~w_t \geq 0. \nonumber
\eea
where the superscript $^{\top}$ denotes transpose. Since $\tau_s$ and $w_t$ are positive operators, their transposed operators are also positive. Similarly to the non-trivial optimization problem discussed in subsection in Eq. (\ref{eq:LEW}), it defines a non-trivial optimization problem to find a decomposition of an EW with positive operators. 

SPAs to EWs can be applied to simplify the task. To this end, we recall the relation in Eq. (\ref{eq:mod}) where the detection condition is given by
\bea
\tr[\widetilde{W}\rho] < \frac{p^*}{d^2}.\nonumber 
\eea
Then, SPAed EWs from Eq. (\ref{eq:relationWW}) in an MDI manner can be constructed as
\bea
\widetilde{W}_{\mathrm{MDI}} = \sum_{s,t}  \gamma_{s,t} \tau_{s}^{\top} \otimes w_{t}^{\top}~~\mathrm{from}~~  \widetilde{W} = \sum_{s,t}  \gamma_{s,t} \tau_{s}  \otimes w_{t}. \nonumber
\eea
This leads to finding a separable decomposition of state $\widetilde{W}$ where recall that $\widetilde{W}$ corresponds to a quantum state. In fact, all EWs can be transformed to a quantum state by admixing some positive operators \cite{Augusiak:2011aa}. This introduces the problem of finding a separable decomposition for a separable CJ state, i.e., $\widetilde{W}$. In fact, it has been shown that EWs obtained from the partial transpose corresponds to the so-called quantum two-design \cite{KalevBae}, that naturally finds a separable decomposition.

\setcounter{footnote}{0}
\section{Conclusion and outlook}
\markboth{\sc Conclusion and outlook}{}
\graphicspath{{conclusion/}}
\label{section:conclusion}

Entanglement is generally a resource for quantum information processing, also one of the resources that make quantum systems to outperform classical ones in information processing. The existence of entangled states is so fundamental that it has the origin in the postulates of quantum theory that, in particular, quantum dynamics is constrained to be a unitary evolution. Consequently, a subsystem dynamics is characterized by positive and CP maps, whereas positive but not CP maps cannot be reduced from unitary evolution. Interestingly, positive but non-CP maps are in fact useful to characterize entangled states in the other way around, and lead to practical methods of detecting entangled states in both theory and experiment. Since entanglement is a resource and thus its detection is necessary for quantum information processing, positive but non-CP maps though being non-physical are of both fundamental and practical interest in quantum information theory. Moreover, they have also been of mathematical interest in the context of operator algebra.  

SPA, transforming non-CP maps to quantum channels, presents a simple and systematic way of approximating those non-physical maps with quantum operations. On the way to reaching the application of SPA to entanglement detection, it has deepened our understanding on entanglement theory and quantum information applications. First of all, as the virtue of positive maps, SPA may allow us to develop various ways of detecting entangled states \cite{Horodecki:2002aa, SPAconjecture, Branciard:2013aa}. On the fundamental point of view to the transpose, SPA shows that the optimal approximation with quantum channels coincides to a measure-and-prepare protocol on the symmetric subspace \cite{KalevBae}. This may show that symmetrization of preparation and measurement provides the optimal way of approximating an anti-unitary transformation. Also on the theoretical side, SPA has shown that, once SPA is applied, the resulting maps are entanglement-breaking in numerous cases of optimal positive maps that give the characterization of entangled states \cite{SPAconjecture, Chruscinski:2009aa, Chruscinski:2010aa, Chruscinski:2011aa, Zwolak:2013aa, Zwolak:2014aa, Augusiak:2014aa}. However, this does not hold in general \cite{Ha:2012aa, Stormer:2013aa, Chruscinski:2014ab, Hansen:2015aa}, and improves the understanding on the structure of quantum states and their characterizations, see a review \cite{Shultz:2016aa}, and also the detailed structure of not only the optimality but also other various properties of extremality, atomicity, spanning property, etc. \cite{Lewenstein:2000ab, Chruscinski:2011aa} concerning to the structure of entangled states. Finally, SPA significantly simplifies implementation of those maps approximating non-physical operations, for which proof-of-principle demonstrations for SPAed maps have been shown for cases of the transpose and the partial transpose with photonic systems \cite{Lim:2011aa, PTexpBae, Lim:2012aa}. 

SPA has initiated a fresh angle to view the relation between entanglement and quantum channels. First, the conjecture in Ref. \cite{SPAconjecture} is motivated by the mere observation that, when the domain is reduced to separable states, positive maps are also CP, as it is discussed in Section \ref{section:spatheory}. However, it does not seem that the optimality plays a significant role since i) there exist counterexamples \cite{Ha:2012aa, Stormer:2013aa, Chruscinski:2014ab, Hansen:2015aa} and also, without resort to the optimality, ii) positive maps detecting all entangled isotropic states satisfy the conjecture \cite{Chruscinski:2009aa, Augusiak:2011aa}. It would be interesting to characterize those positive maps that satisfy the conjecture. For such maps, experimental realization is also feasible with present-day technologies. Next is a general link between quantum $t$-design and the structure of POVMs of SPAed maps. For the case of the transpose, it appears that quantum two-design elucidates the structure of POVMs of the SPAed map; in other words, the anti-unitary transformation is optimally approximated by the symmetrization. Since both of quantum design and SPA is fundamental, their relation may sharpen the understanding on entanglement. Finally, SPA has shown various connections of fundamental and application aspects of quantum information processing. It is of both fundamental and practical interest to devise further applications of SPA and also to realize SPAed maps in various physical systems.

\section*{Acknowledgement}

The author is grateful to A. Acin, M. Almeida, R. Augusiak, J. Korbicz, M. Lewenstein and J. Tura for collaborations and discussions that have always led to interesting directions, in particular to finding the usefulness and significance of the relation between SPA and entanglement-breaking channels. The author thanks A. Kalev for collaborations on quantum two-design, D. Chruscinski for discussions and comments on positive maps and Choi-Jamiolkowski isomorphism, and K.-C. Ha and S.-H. Kye for discussions and related problems in Operator Algebra. Collaboration with experimentalists H.-T. Lim, Y.-S. Ra, Y.-S. Kim, and Y.-H. Kim taking their resources and efforts for the implementation is greatly acknowledged. 

This work is supported Institute for Information \& communications Technology Promotion(IITP) grant funded by the Korea government(MSIP) (No.R0190-17-2028, PSQKD), the KIST Institutional Program (Progect No. 2E26680-17-P025), and the People Programme (Marie Curie Actions) of the European Union Seventh Framework Programme (FP7/2007-2013) under REA grant agreement N. 609305.






\end{document}